\documentclass[twocolumn]{aastex6}
\usepackage{natbib}
\usepackage{color}
\usepackage{hyperref} 
\usepackage{booktabs}

\def    \apjl  		{\rm {ApJL}}
\def    \apj  		{\rm {ApJ}}
\def    \mnras  	{\rm {MNRAS}}
\def    \araa  		{\rm {ARA\& A}}
\def    \apjl  		{\rm {ApJL}}

\def	\cm		{\,{\rm {cm}}}
\def	\K		{\,{\rm K}}

\def \bea {\begin{eqnarray}}
\def \ena {\end{eqnarray}}

\def	\cm	{\,{\rm cm}}

\def	\erg	{\,{\rm erg}}

\def	\H	{{\rm H}}

\usepackage{amsmath}

\begin{document}
\shorttitle{Polarized dust emission from $\rho$ Ophiuchi A}
\shortauthors{Tram et al.}
\title{Understanding polarized dust emission from $\rho$ Ophiuchi A in light of grain alignment and disruption by radiative torques}

\author{Le Ngoc Tram\altaffilmark{1,2}, Thiem Hoang\altaffilmark{3,}\altaffilmark{4}, Hyeseung Lee\altaffilmark{3}, Fabio P. Santos\altaffilmark{5}, Archana Soam \altaffilmark{1}, Pierre Lesaffre\altaffilmark{6,7}, Antoine Gusdorf\altaffilmark{6,7} and William T. Reach\altaffilmark{1}}

\affil{$^1$ Stratospheric Observatory for Infrared Astronomy, Universities Space Research Association, NASA Ames Research Center, MS 232-11, Moffett Field, 94035 CA, USA; \href{mailto:ngoctram.le@nasa.gov}{ngoctram.le@nasa.gov}}
\affil{$^2$ University of Science and Technology of Hanoi, VAST, 18 Hoang Quoc Viet, Vietnam}
\affil{$^3$ Korea Astronomy and Space Science Institute, Daejeon 34055, South Korea}
\affil{$^4$ Korea University of Science and Technology, 217 Gajeong-ro, Yuseong-gu, Daejeon, 34113, South Korea}
\affil{$^5$ Max-Planck-Institute for Astronomy, K\"{o}nigstuhl 17, D-69117 Heidelberg, Germany}
\affil{$^6$ Laboratoire de Physique de l'\'Ecole normal sup\'erieur, ENS, Universit\'e PSL, CNRS, Sorbonne Universit\'e, Universit\'e de Paris, France}
\affil{$^7$ Observatoire de Paris, PSL University, Sorbonne Universit\'e, LERMA, 75014, Paris, France}
\begin{abstract}
The alignment of dust grains with the ambient magnetic field produces polarization of starlight as well as thermal dust emission. Using the archival SOFIA/HAWC+ polarimetric data observed toward $\rho$ Ophiuchus (Oph) A cloud hosted by a B association at 89 $\mu$m and 154 $\mu$m, we find that the fractional polarization of thermal dust emission first increases with the grain temperature and then decreases once the grain temperature exceeds $\simeq 25-32\K$. The latter trend differs from the prediction of the popular RAdiative Torques (RATs) alignment theory which implies a monotonic increase of the polarization fraction with the grain temperature. We perform numerical modeling of polarized dust emission for the $\rho$ Oph-A cloud and calculate the degree of dust polarization by simultaneously considering the dust grain alignment and rotational disruption by RATs. Our modeling results could successfully reproduce both the rising and declining trends of the observational data. Moreover, we show that the alignment of only silicate grains or a mixture of silicate-carbon grains within a composite structure can reproduce the observational trends, assuming that all dust grains follow a power-law size distribution. Although there are a number of simplifications and limitations to our modeling, our results suggest grains in $\rho$ Oph-A cloud have a composite structure, and the grain size distribution has steeper slope than the standard size distribution for the interstellar medium. Combination of SOFIA/HAWC+ data with JCMT observations 450~$\mu$m and 850~$\mu$m would be useful to test the proposed scenario based on grain alignment and disruption by RATs.
\end{abstract}

\section{Introduction\label{sec:intro}}
The magnetic field is believed to play an essential role in various astrophysical phenomena, including the formation of stars and planets (\citealt{2012ARA&A..50...29C}). The alignment of dust grains with the magnetic field induces polarization of starlight and of thermal dust emission. The polarization vectors of starlight are parallel to the magnetic field, while those of thermal dust are perpendicular to the magnetic field. Thus, dust polarization has become a popular technique to constrain the magnetic field direction and strength (\citealt{2007JQSRT.106..225L}; \citealt{2015ARA&A..53..501A}).

Observations have reported an anti-correlation trend of the fractional polarization of thermal dust emission with the column density of the gas in molecular clouds (e.g., \citealt{1998ApJ...499L..93A}; \citealt{2008ApJ...674..304W}; \citealt{2015A&A...576A.105P}; \citealt{2016ApJ...824..134F}; \citealt{2017ApJ...837..161S,2019ApJ...882..113S}; \citealt{2018arXiv180706212P}; \citealt{2019ApJ...872..187C}). This trend is explained by the loss of grain alignment toward dense cores (\citealt{2008ApJ...674..304W}) or by the turbulent structure of magnetic field within the scale of the beam size (see \citealt{2015psps.book..147J}; \citealt{2015A&A...576A.105P}).   

A popular theory describing grain alignment is RAdiative Torques (hereafter referred to as RATs) (\citealt{2007MNRAS.378..910L}; see \citealt{2007JQSRT.106..225L}; \citealt{2015ARA&A..53..501A} for reviews). One of the key predictions of the RAT theory is that the polarization degree correlates with the intensity of the radiation field (or equivalently dust temperature $T_{\rm d}$). This prediction was numerically demonstrated by \cite{2020ApJ...896...44L}. However, observations revealed that the dust polarization degree does not always increase with $T_{\rm d}$. For example, \cite{2018arXiv180706212P} showed that the 40' spatial resolution polarization degree at 850 $\mu$m, measured by the \textit{Planck} satellite toward four molecular regions, including Aquila Rift, Cham-Musca, Orion, and Ophiuchus in the Gould belt cloud, decreases for $T_{\rm d}>19~\K$ (see their Figure 18). Additionally, far-Infrared polarimetric data observed by the High-resolution Airborne Wide band Camera Plus (HAWC+) instrument (\citealt{2018JAI.....740008H}) onboard the Stratosphere Observatory for Infrared Astronomy (SOFIA) toward the molecular cloud Ophiuchus A (\citealt{2019ApJ...882..113S}) at 89 $\mu$m (7.8'' spatial resolution) and 154 $\mu$m (13.6'' spatial resolution) also reported the decrease of the polarization degree for $T_{\rm d}>32\K$ (see Section \ref{sec:obs} below). These observational features are challenging the popular RAT alignment theory.     

Dust grain-size distribution is an important parameter when it comes to interpreting the polarization of dust. The grain size distribution is expected to evolve from the diffuse interstellar medium (ISM) to dense molecular clouds (MCs) due to grain accretion of gas species and grain-grain collisions (\citealt{2013MNRAS.434L..70H}). Recently, \cite{2019NatAs...3..766H} discovered that a large grain exposed to a strong radiation field could be disrupted into small fragments due to centrifugal stress induced by suprathermal rotation by RATs. This effect is termed Radiative Torques Disruption (RATD) (see \citealt{2020arXiv200616084H} for a review). Since RATs are stronger for larger grains
(\citealt{2007MNRAS.378..910L}; \citealt{2008MNRAS.388..117H}), RATD
is more efficient for large grains than smaller ones. As shown
in \cite{2019ApJ...876...13H}, the RATD mechanism is much faster than
grain shattering and thus determines the upper cutoff of the
grain size distribution in the ISM.

\cite{2020ApJ...896...44L} carried out numerical modeling of multi-wavelength polarization of thermal dust emission from aligned grains by RATs. They show that the polarization degree at 850 $\mu$m first increases with increasing dust temperature. However, when RATD is accounted for, they find that the polarization degree decreases for higher dust temperature, which is different from classical RATs prediction. The level of the decline is found to depend on the tensile strength, which is determined by the internal structure of dust grains (\citealt{2019ApJ...876...13H}). Interestingly, accounting for RATD, the model predicts the same $P(\%)-T_{\rm d}$ trend as reported by \textit{Planck} data (\citealt{2018arXiv180706212P}) at the same wavelength as mentioned above. The success of the joint effect of RAT alignment and RATD in explaining {\it Planck} data motivates us to use this approach to better interpret the SOFIA/HAWC+ data.

Coming back to the SOFIA/HAWC+ observation toward $\rho$ Oph-A at band C (89 $\mu$m) and D (154 $\mu$m) as mentioned above, \cite{2019ApJ...882..113S} mainly studied the variation of the ratio of the polarization degree ($P_{\rm D}(\%)/P_{\rm C}(\%)$) with respect to the dust temperature, which is opposed to the polarization degree studies. Furthermore, the authors showed that classical RATs mechanism was able to explain the increasing (e.g., positive) part of the ratio curve and discarded the decreasing (negative) part (see their Figure 6d). In this study, we will: (1) use this dataset to show the correlation to the polarization degree itself to dust temperature; and (2) extend the improved polarized thermal dust model introduced by \cite{2020ApJ...896...44L} to interpret these SOFIA/HAWC+ observational trends.         

This paper is structured as follows. We present the archival SOFIA/HAWC+ data from $\rho$ Oph-A at $89~\mu$m and $154~\mu$m observed by SOFIA/HAWC+ in Section \ref{sec:obs}. We describe our modeling method of polarized thermal dust emission by aligned grains in Section \ref{sec:model}. In Section \ref{sec:compare}, we compare our numerical results obtained for $\rho$ Oph-A cloud with observational data. A summary of our findings and conclusions are presented in Section \ref{sec:discussion}. 

\section{Observations toward $\rho$ Oph-A} \label{sec:obs}
\begin{figure*}
    \centering
    \includegraphics[width=0.45\textwidth]{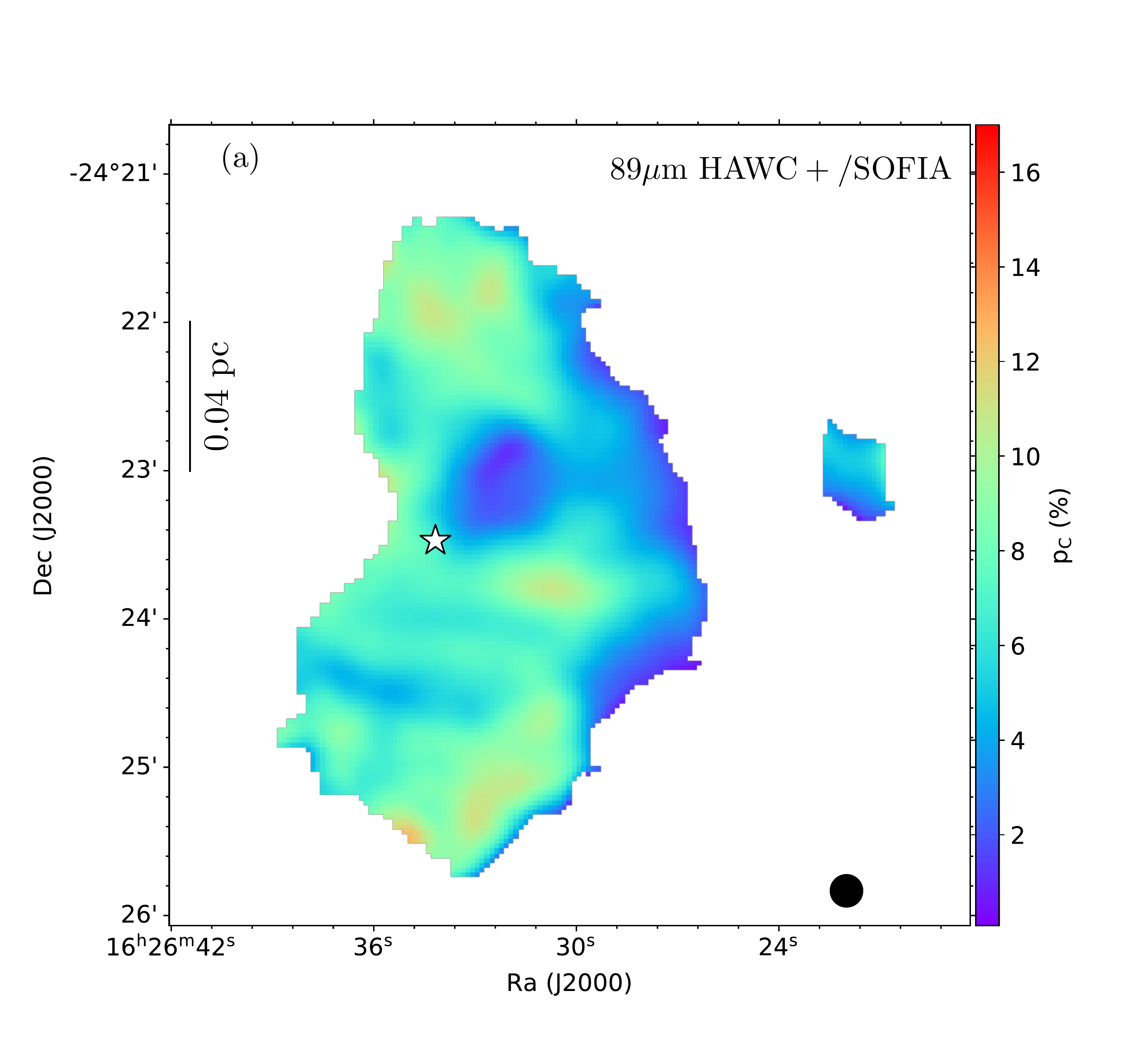}
    \includegraphics[width=0.45\textwidth]{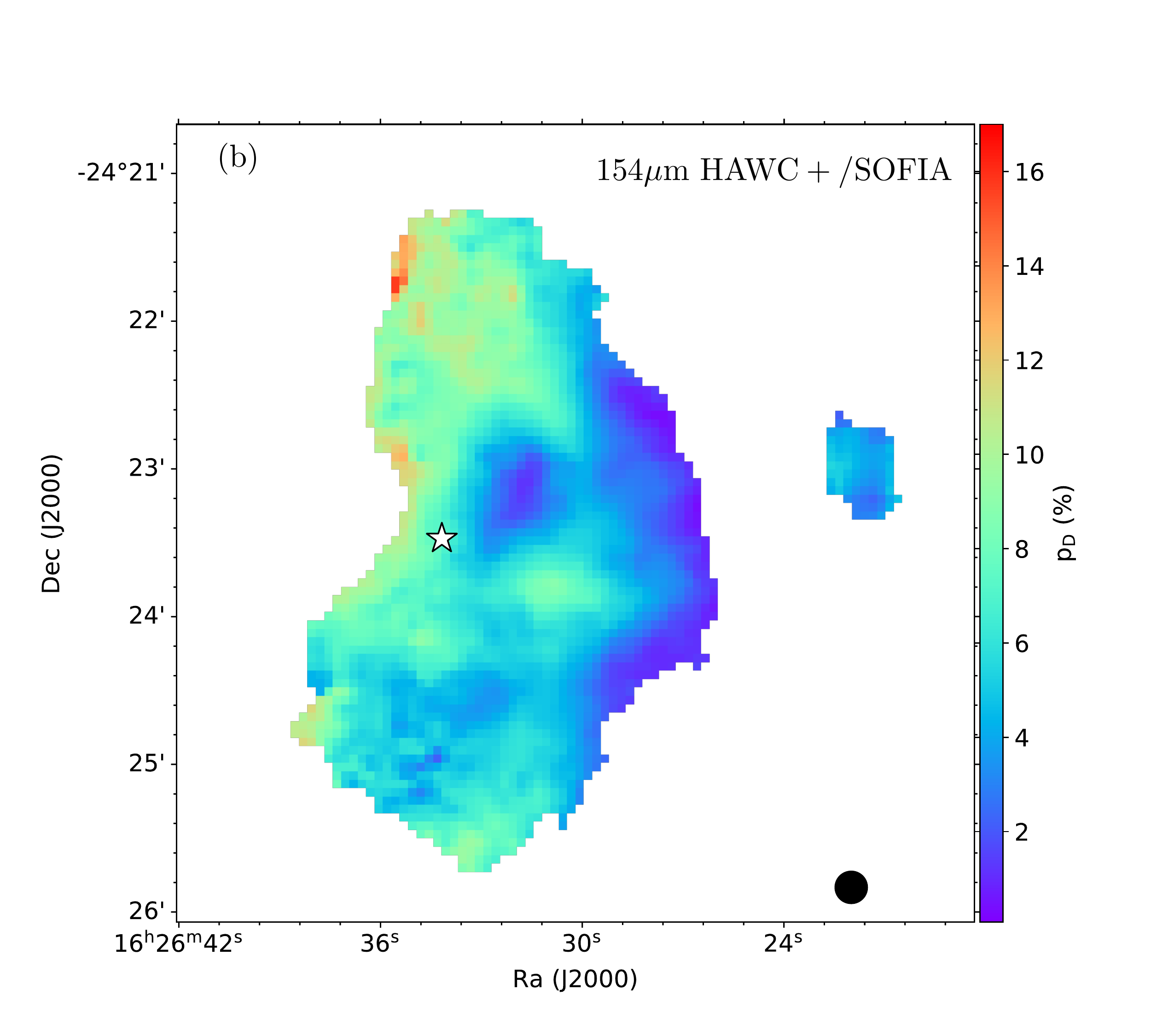}
    \includegraphics[width=0.45\textwidth]{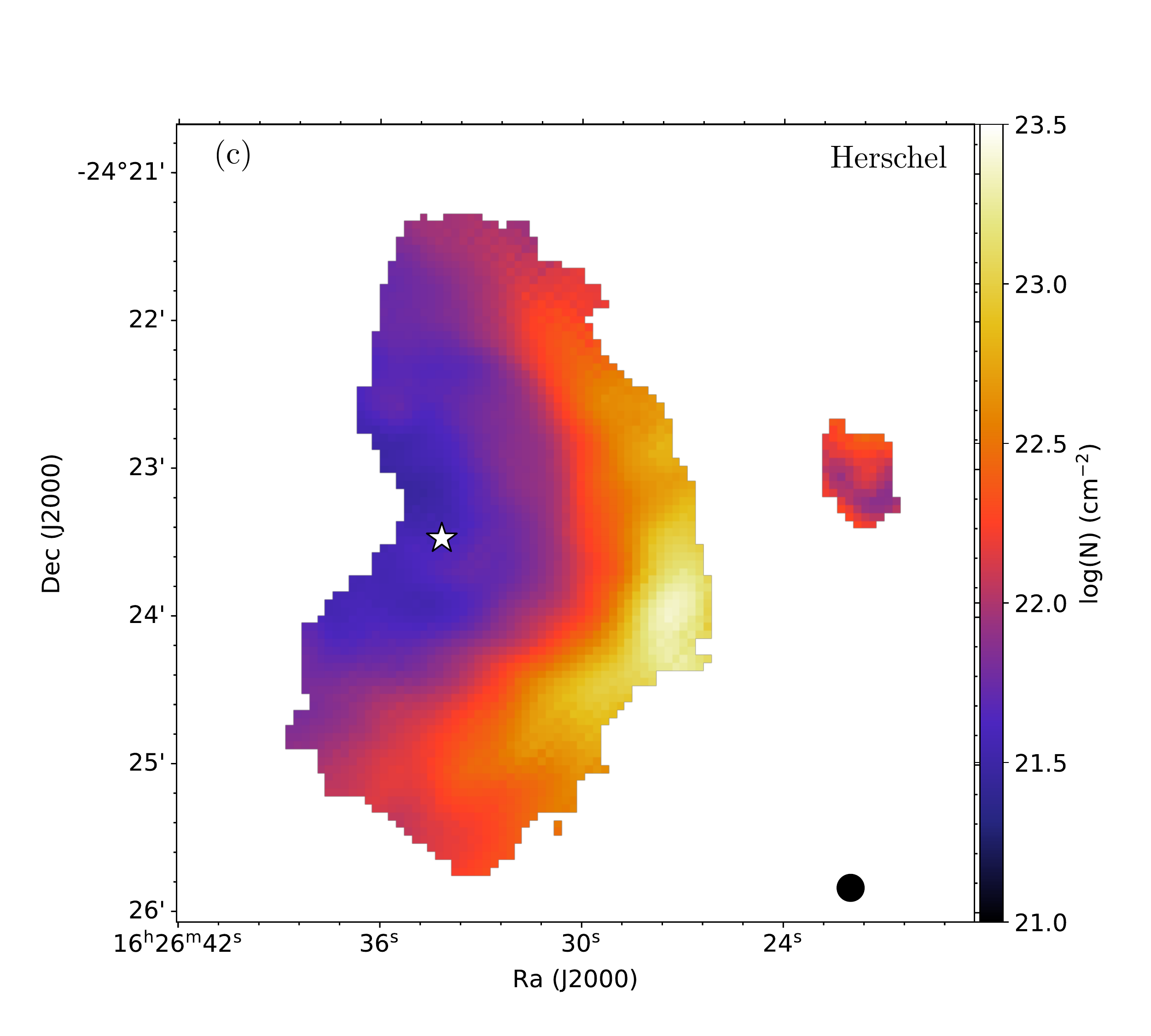}
    \includegraphics[width=0.45\textwidth]{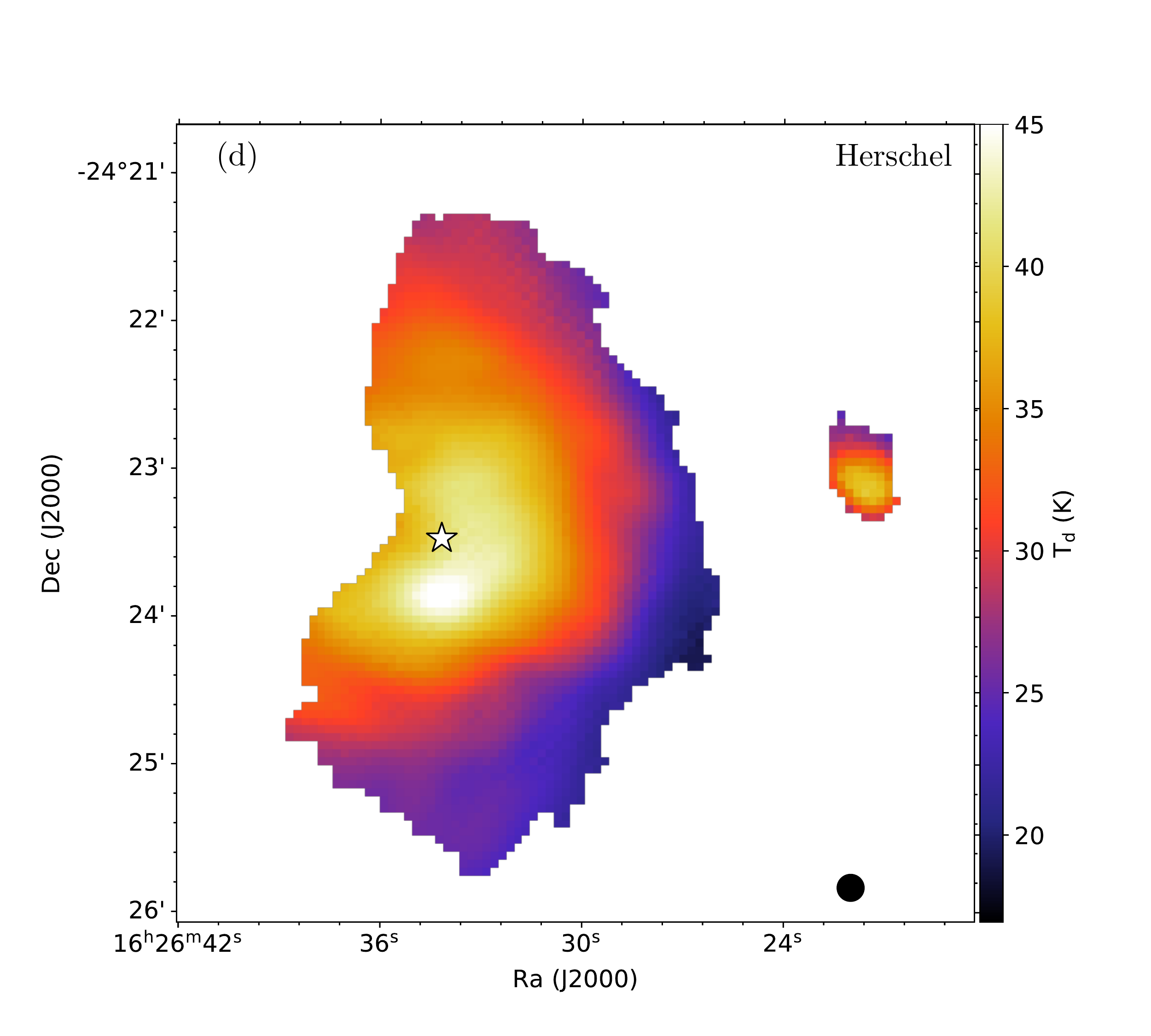}
    \caption{Maps of polarization degree and dust temperature of $\rho$ Oph-A. (a) and (b) The polarization degrees measured by bands C and D of HAWC+/SOFIA in the actually resolutions. (c) and (d) The maps of the H$_{2}$ column density ($N$) and the dust temperature ($T_{\rm d}$) derived from 70, 100, 160$\,\mu$m PACS/\textit{Herschel} data. The star symbol locates the position of Oph S1. The black filled circles show the beam size. The physical scale is derived from 140 pc of distance.}
    \label{fig:polametric_maps}
\end{figure*}
$\rho$ Oph-A is a molecular cloud in one of the closest dark cloud complex and star-forming region $\rho$ Ophiuchi. Distance to this complex is reported to be $\sim$ 120--160 pc \citep{1981A&A....99..346C, 1989A&A...216...44D, 1998A&A...338..897K, 2004AJ....127.1029R, 2008ApJ...675L..29L, 2008A&A...480..785L, 2008AN....329...10M, 2008ApJ...679..512S, 2017ApJ...834..141O}. This region is significantly influenced by high energy radiation from a high-mass Oph-S1 star, which is a B association star (\citealt{1977AJ.....82..198V}; \citealt{1988ApJ...335..940A}; \citealt{1989ApJ...338..902L, 1989ApJ...338..925L};\citealt{2003PASJ...55..981H}). Among several dark clouds cores in $\rho$ Ophiuchi, Oph-A is identified as one of the warmest cores compared to Oph-B and C regions. Several studies i.e. $Herschel$, $Spitzer$, and James Clarke Maxwell Telescope (JCMT) Gould belt surveys (\citealt{2010A&A...518L.102A}; \citealt{2009ApJS..181..321E}; \citealt{2007PASP..119..855W}) include this region for various investigations on dust and gas properties.
This cloud complex is also widely studied in multi-wavelength imaging and polarimetry. Recent attempts were made to map magnetic fields in Oph-A region using near-IR and sub-mm polarization measurements by \cite{2015ApJS..220...17K, 2018ApJ...859....4K} and far-IR by \cite{2019ApJ...882..113S}, respectively. Oph-A is one of the best laboratory to understand the multi-band dust polarization in context of high energy radiation giving opportunity to investigate RAT in detail. 

\begin{figure*}
    \centering
    \includegraphics[width=0.95\textwidth]{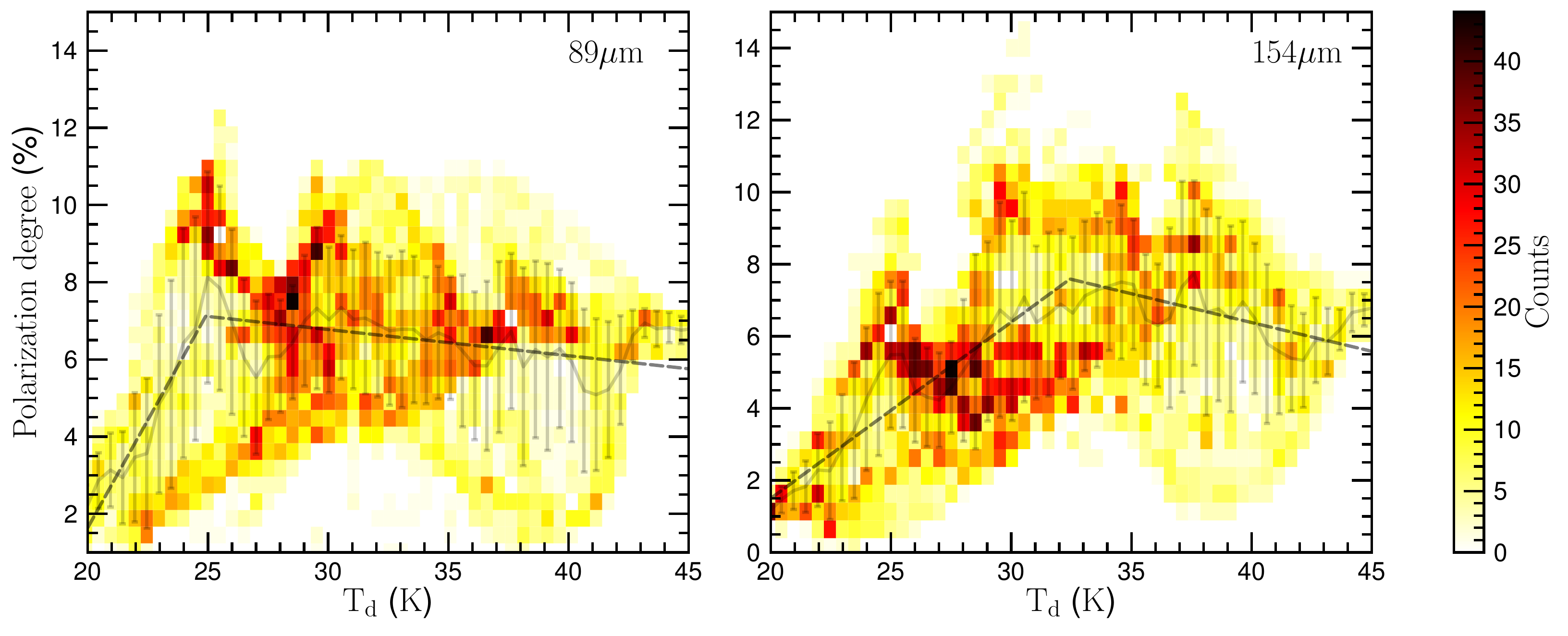}
    \caption{2D histogram of the dust polarization degree and dust temperature for 89 $\mu$m (left panel) and 154 $\mu$m (right panel). This diagram is made of 52 bins and the gray lines show the binning weighted-mean of the data and the error bars represent the standard deviation within the bin. The black dashed lines show the best fit of the piecewise linear function to the data. The maps of the dust temperature and polarization at 89 $\mu$m are smoothed to $13''.6$ of FWHM.}
    \label{fig:hist2d_maps}
\end{figure*}
\subsection{Polarization maps} \label{sec:pol_map}
In this work, we use the archival FIR polarimetric data observed by SOFIA/HAWC+. These data sets are introduced in \cite{2019ApJ...882..113S}. The observations were made in 2017 using two bands of HAWC+ instrument, namely C (89 $\mu m$) and D (154 $\mu$m). The angular resolutions are $7''.8$ and $13''.6$, respectively. The polarization degree maps in these bands are shown in Figure \ref{fig:polametric_maps}(a,b)\footnote{We smoothed the $7''.8$ band C and the $11''.4$ dust temperature maps to $13''.6$ band D resolution using the python-package \textsc{Gaussian2DKernel}.}. We select the common sky positions in which data is detected in both bands. The local polarization degree varies significantly across the $\rho$ Oph-A cloud, in which the median value is 7.5$\%$ in band C and 5.0$\%$ in band D as discussed in \cite{2019ApJ...882..113S}. Figure \ref{fig:polametric_maps}(a,b) shows a "tight" spatial correlation in polarization degree in two bands except at the southernmost area ($T_{\rm d} \simeq 25\K$), where the data at 89$\,\mu$m is more polarized than at 154$\,\mu$m. The reason for this such a difference is beyond this work's scope because we do not have enough information to investigate it quantitatively. However, a possible explanation could be that there is a warmer outer component (with the local temperature larger than $25\K$) and a colder inner component (with the local temperature smaller than $25\K$) along the line-of-sight (LOS) as proposed in \citealt{2019ApJ...882..113S}. In the warmer component, which favors enhancing the shorter wavelength polarization (band C), grains are well exposed to radiation. Therefore the alignment is efficient, causing a larger value of $P_{\rm C}(\%)$. On the contrary, the colder component along the LOS favors emission at longer wavelengths (band D); however, the grain alignment is less effective due to the shielding. Therefore, the value of $P_{\rm D}(\%)$ become smaller. A star symbol locates the high-mass star Oph S1.

\subsection{Map of dust temperature and gas density}
We adopt the dust temperature ($T_{\rm d}$) and the gas column density ($N$) maps of \cite{2019ApJ...882..113S}. These maps were generated by a fit of the modified thermal spectral energy distribution (SED) to each pixel using 70, 100, and 160$\,\mu$m from \textit{Herschel/PACS} data (\citealt{2010A&A...518L...2P}) with the fixed exponential index of the dust opacity 1.6. Figure \ref{fig:polametric_maps}(c,d) shows the gas density and dust temperature maps in the same regions that HAWC+ detected data. The high-mass star Oph S1 warms up the surrounding environment, causing a large temperature gradient, i.e., from $\simeq 45\K$ near Oph S1 down to $\simeq 20\K$ at the edge of the cloud. On the contrary, the gas is densest at the edge of the map and radially diffuses backward to Oph S1.  

\subsection{Dust polarization and temperature}
Figure \ref{fig:hist2d_maps} shows the 2D-histogram dust polarization degree in band C (left panel) and D (right panel) to dust temperature made of 52 bins. They share the same feature (1) the polarization degree increases as the dust temperature increases up to $T_{\rm d} \simeq T_{\rm crit}$ (i.e., positive slope region) and (2) the polarization degree decreases for higher dust temperature (i.e., negative slope region). In the positive slope, the polarization degree in band C is higher than band D, while it is lower in the negative slope, which was also showed by these fractional polarization ratio by Figure 6d in \citealt{2019ApJ...882..113S}). In other words, the polarization degree at a shorter wavelength (89 $\mu$m) is higher than at the longer wavelength (154 $\mu$m) in the denser region (i.e., at the edge of the polarimetric map) while it is the opposite in the less dense region (close central star) in $\rho$ Oph-A.  
Using the RATs theory, the spherical model of \cite{2019ApJ...882..113S} could explain the increase (decrease) of the $P_{\rm D}/P_{\rm C}$ ratio with respect to dust temperature (gas column density) in the dense ($N>10^{21.75}\rm cm^{-2}$) and cold region ($T_{\rm d} \leq 32-34\K$) (see their Figure 6). However, this model could not explain the observational trend in the more diffuse and hotter region. We fitted the data with a piecewise linear function\footnote{We used the python-package \textsc{pwlf} (piecewise linear fitting).}. The best fits (i.e., black dashed lines) show that the transition takes place at $T_{\rm crit} \simeq 25\K$ for band C, and $\simeq 32\K$ for band D. In the statistical point of view, the reason for low $T_{\rm crit}$ in band C is that there is an excess of the polarization degree at $T_{\rm d} \simeq 25\K$ as mentioned in Section \ref{sec:pol_map}, which makes the piecewise linear fit peak toward this value of $T_{\rm d}$. In the framework of our theoretical point of view, however, the transition from positive to negative slope at $\simeq 25\K$ is unlikely physical because the tensile strength of grains must be extremely small (see Section \ref{sec:compare}). In addition, the polarization ratio $P_{\rm D}/P_{\rm C}$ changes its slope at $T_{\rm d}\simeq 32-34\K$ as shown in \cite{2019ApJ...882..113S}.

\section{Modelling thermal dust polarization} \label{sec:model}
\begin{figure}
    \centering
    \includegraphics[width=0.5\textwidth]{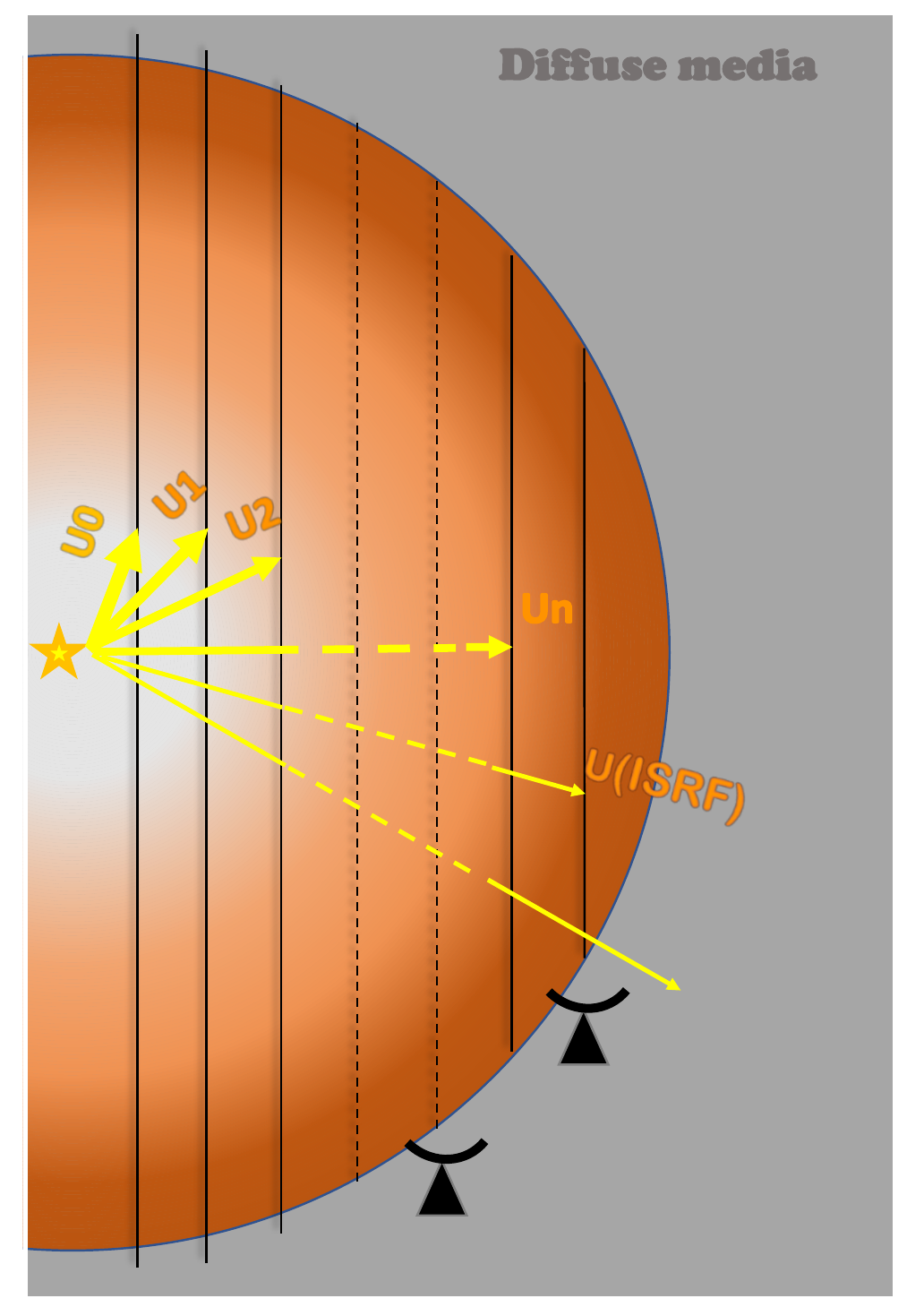}
    \caption{Schematic illustration of the model. A molecular cloud is irradiated by the central star, so that the radiation strength (in equivalent to dust temperature) decreases from $U_{0}$ to $U_{\rm ISRF}$ The color gradient indicates the increase of the gas density from the central star.}
    \label{fig:model_schem}
\end{figure}

The multi-wavelength polarization model of the thermal dust emission is described in detail in \cite{2020ApJ...896...44L}. The schematic of the model is illustrated in Figure \ref{fig:model_schem}. The radiative source (e.g., a O/B star) is denoted by a star symbol surrounded by a cloud. The radiative strength ($U$) gets smaller at further in the cloud. As follows, we describe the model to calculate the polarization of thermal dust emission from this cloud.
\subsection{Fractional polarization of thermal emission}
Dust grains are heated by the radiation and re-emit in the thermal range. The fractional polarization of the thermal dust emission is the ratio of the polarized intensity ($I_{\rm pol}$) to the total emission intensity ($I_{\rm em}$), which yields 
\bea \label{eq:pol_degree}
    P(\%) = 100\times \frac{I_{\rm pol}}{I_{\rm em}}.
\ena

Assuming a dust environment containing carbonaceous and silicate grains, the total emission intensity is given by
\bea
    \frac{I_{\rm em}(\lambda)}{N_{\H}} = \sum_{j=\rm sil,car} &&\int^{a_{\rm max}}_{a_{\rm min}} Q_{\rm ext}\pi a^{2} \nonumber\\
    &\times&\int dT B_{\lambda}(T_{\rm d})\frac{dP}{dT}\frac{1}{n_{\rm H}}\frac{dn_{j}}{da}da.~~~
\ena

If silicate and carbon are separated populations, then as paramagnetic grains, silicates can align with the ambient magnetic field, while carbon grains cannot (\citealt{2016ApJ...831..159H}). Thus, the polarized intensity resulting from its alignment is given by
\bea
    \frac{I_{\rm pol}(\lambda)}{N_{\rm H}}= &&\int^{a_{\rm max}}_{a_{\rm min}} f(a)Q_{\rm pol}\pi a^{2} \nonumber\\
    &\times&\int dT B_{\lambda}(T_{\rm d})\frac{dP}{dT}\frac{1}{n_{\rm H}}\frac{dn_{sil}}{da}da, ~~~
\ena
where $B_{\lambda}(T_{\rm d})$ is the black-body radiation at dust temperature $T_{\rm d}$, $dP/dT$ is the distribution of dust temperature, $f(a)$ is the alignment function, Q$_{ext}$ is the extinction coefficient, $Q_{\rm pol}$ is the polization coefficient, $dn/da$ is the grain-size distribution. The dust temperature distribution depends on the grain size and radiation strength, which is computed by the DustEM code (\citealt{2011A&A...525A.103C}, see e.g., Figure 8 in \citealt{2020ApJ...896...44L}). The extinction and polarization coefficients are computed by the DDSCAT model (\citealt{1994JOSAA..11.1491D, 2008JOSAA..25.2693D}; \citealt{2012OExpr..20.1247F}) for a prolate spheroidal grain shape with an axial ratio of 1/3. 

If silicate and carbon grains are mixed together (e.g., \citealt{2013A&A...558A..62J}), which may exist in dense clouds due to many cycles of photo-processing, coagulation, shattering, accretion, and erosion, carbon grains could be aligned with the ambient magnetic field and its thermal emission could be polarized. For a simplest case, assuming these grain populations have the same alignment parameters (i.e., $a_{\rm align}$, $f(a)$), the total polarized intensity is
\bea
    \frac{I_{\rm pol}(\lambda)}{N_{\rm H}} = \sum_{j=\rm sil,car} &&\int^{a_{\rm max}}_{a_{\rm min}} f(a)Q_{\rm pol}\pi a^{2} \nonumber\\
    &\times&\int dT B_{\lambda}(T_{\rm d})\frac{dP}{dT}\frac{1}{n_{\rm H}}\frac{dn_{j}}{da}da.
\ena

\subsection{Radiative torques disruption and grain-size distribution} \label{sec:RATD}
Let us consider a radiation field with the energy density of $u_{\rm rad} (\erg \cm^{-3})$, the mean wavelength of $\bar{\lambda}$ and an anisotropy degree of $\gamma$. Its strength is defined by a dimensionless $U=u_{\rm rad}/u_{\rm ISRF}$, where $u_{\rm ISRF}=8.64\times 10^{-13}\erg\cm^{-3}$ is the radiation energy density of the interstellar radiation field (ISRF) in the solar neighborhood (\citealt{1983A&A...128..212M}). This radiation field can spin a dust grain of size $a$ and density $\rho$ up to the rotational rate\footnote{Note that $\omega_{\rm RAT}/\omega_{\rm T} \sim 1/(1+F_{\rm IR})$, not $\sim (1+F_{\rm IR})$ as the typo in \cite{2020ApJ...896...44L}, Equation (3).}
\bea \label{eq:omega_RAT}
    \frac{\omega_{\rm RAT}}{\omega_{\rm T}} \simeq  &&2.9\times 10^{2}  \hat{\rho}^{0.5} \gamma a^{3.2}_{-5} U\left(\frac{10^{3} \rm{cm^{-3}}}{n_{\H}}\right)\left(\frac{\bar{\lambda}}{0.5 \rm{\mu m}}\right)^{-1.7} \nonumber\\
    &\times& \left(\frac{20 \K}{T_{\rm gas}}\right)\left(\frac{1}{1+F_{\rm IR}}\right),
\ena
where $a_{-5}=a/(10^{-5} \rm cm)$, $\hat{\rho}=\rho/(3 \rm g cm^{-3})$, $n_{\H}, T_{\rm gas}$ are the gas density and temperature. $\omega_{\rm T}=(k_{\rm B}T_{\rm gas}/I)^{0.5}$ is the thermal angular velocity with $I=8\pi \rho a^{5}/15$ the inertia moment of grain. A rotating grain is damped by gas collisions and IR emission (see \citealt{2019ApJ...876...13H}). The dimensionless parameter ($F_{\rm IR}$) that describes the ratio of the IR damping to collisional damping \footnote{The factor is corrected to be 0.4 from Equation (4) in \cite{2020ApJ...896...44L}.} is defined as 
\bea
    F_{\rm IR} \simeq 0.4\left(\frac{U^{2/3}}{a_{-5}}\right)\left(\frac{30 \rm{cm^{-3}}}{n_{\H}}\right)\left(\frac{100\K}{T_{\rm gas}}\right)^{1/2}.
\ena

A grain rotating at rotational velocity $\omega$ results in a tensile stress $S=\rho \omega^{2}a^{2}/4$ on the materials making up the grain. Thus, the maximum rotational velocity that a grain can withstand is:
\bea \label{eq:omega_crit}
    \omega_{\rm crit} = \frac{2}{a}\left(\frac{S_{\rm max}}{\rho}\right)^{1/2} \simeq \frac{3.6\times 10^{8}}{a_{-5}}S^{1/2}_{\rm max,7}\hat{\rho}^{-1/2},
\ena
where $S_{\rm max,7}=S_{\rm max}/(10^{7} \rm erg \cm^{-3})$. 

One can see from Equation (\ref{eq:omega_RAT}) that the stronger the radiation field and the larger the grain size, the faster the rotation of the grain. A strong radiation field can thus generate such a fast rotation that the induced stress on large grains can result in a spontaneous disruption. This disruption mechanism is named as RATD and discovered by \cite{2019NatAs...3..766H}. From Equations (\ref{eq:omega_RAT}) and (\ref{eq:omega_crit}), we can derive the critical size above which grains are disrupted:
\bea
    \left(\frac{a_{\rm disr}}{0.1\,\rm \mu m}\right)^{2.7} \simeq 5.1\gamma^{-1}_{0.1}U^{-1/3}\bar{\lambda}^{1.7}_{0.5}S^{1/2}_{\rm max,7},
\ena
where $\bar{\lambda}_{0.5} = \bar{\lambda}/(0.5\,\mu \rm m)$. 

Dust grains are disrupted efficiently (for $a$ greater than $a_{\rm disr}$) in stronger radiation fields. The disruption of dust grain by RATD can modify the grain-size distribution. Since only the largest grains are affected by the RATD mechanism, RATD determines the upper limit of the size distribution. The disruption is thus expected to enhance more smaller grains, resulting in a steeper grain size distribution than in the standard ISM. In the particular case of $\rho$ Oph-A cloud, \cite{2015A&A...578A.131L} furthermore showed that the grain size distribution experiences a varying power index across the cloud. In this work, we adopt a power-law grain size distribution assumption for both the original large grains and the smaller grains produced by disruption, with a power-law index $\beta$:
\bea
    \frac{1}{n_{\H}}\frac{dn_{\rm sil,car}}{da}=C_{\rm sil,car}a^{\beta} \ \ \ \rm{(a_{\rm min}\leq a \leq a_{\rm max})},
\ena
where $C_{\rm sil}$ and $C_{\rm car}$ are the normalization constants for silicate and carbonaceous grains, respectively. The smallest grain size is chosen as $a_{\rm min}=10~\AA$, while the maximum size is constrained by the RATD mechanism (i.e., $a_{\rm max}=a_{\rm disr}$). The normalization constants are determined through the dust-to-gas mass ratio $M_{\rm d/g}$ (see \citealt{2020arXiv200906958C}; \citealt{2020ApJ...893..138T}) as
\bea 
    \sum_{\rm j=sil,car} C_{j}\rho_{j} &=& \frac{(4+\beta)M_{\rm d/g} m_{\rm gas}}{\frac{4}{3}\pi (a^{4+\beta}_{\rm max}-a^{4+\beta}_{\rm min})} ~~~~~~~\rm{for\ \beta \neq -4} \\ \nonumber
    \sum_{\rm j=sil,car} C_{j}\rho_{j} &=& \frac{M_{\rm d/g}m_{\rm gas}}{\frac{4}{3}\pi(\ln a_{\rm max} - \ln a_{\rm min})} ~~~\rm{for\ \beta = -4}.
\ena
where $C_{\rm sil}/C_{\rm car}$ is adopted as 1.12 (\citealt{1984ApJ...285...89D}), and the dust-to-mass ratio $M_{\rm d/g}$ is fixed as 0.01 throughout this work. The latter assumption is close to what is derived from X-ray observations $\simeq 0.011 - 0.0125$ (\citealt{2003A&A...408..581V}) or gas tracers $\simeq 0.0114$ (\citealt{2015A&A...578A.131L}). 

\subsection{Grain alignment by RATs}
An anisotropic radiation field can align dust grains
via the RATs mechanism (see \citealt{2007JQSRT.106..225L}; \citealt{2015ARA&A..53..501A} for reviews). In the unified theory
of RATs alignment, grains are first spun-up to suprathermal rotation and then driven to be aligned with the ambient magnetic fields by superparamagnetic relaxation within grains having iron inclusions (\citealt{2016ApJ...831..159H}). Therefore, grains are only efficiently aligned
when they can rotate suprathermally. This aligned grain size ($a_{\rm align}$) is determined by the following condition $\omega_{\rm RAT}(a_{\rm align}) = 3\omega_{T}$ as in \cite{2008MNRAS.388..117H}. From Equation \ref{eq:omega_RAT}, we have:
\bea
    a_{\rm align} \simeq &&0.024\hat{\rho}^{-5/32} \gamma^{-5/16} U^{-5/16} \left(\frac{10^{3} \cm^{-3}}{n_{\H}}\right)^{-5/16} \\ \nonumber
    &&\times \left(\frac{\bar{\lambda}}{0.5\rm \mu m}\right)^{17/32} \left(\frac{20\K}{T_{\rm gas}}\right)^{-5/16}\left(\frac{1}{1+F_{\rm IR}}\right)^{-5/16} ~\rm{\mu m},
\ena 
which implies $a_{\rm align} \sim 0.02\,\mu$m for a dense ISM with $\gamma=1.0$, $U=1$, and $\bar{\lambda}=0.3\,\mu$m. In this work, we adopt the alignment function as in \cite{2020ApJ...896...44L}:
\bea \label{eq:fa}
    f(a)=f_{\rm min}+(f_{\rm max}-f_{\rm min})\left\{1-\exp{\left[-\left(\frac{0.5a}{a_{\rm align}}\right)^{3}\right]}\right\}.~~~
\ena
For those grains with $a\ll a_{\rm align}$, the alignment is minimum as $f_{\rm min}=10^{-3}$, while the alignment degree gets maximum $f_{\rm max}$ for $a\gg a_{\rm align}$. This parametric function agrees the results obtained from inverse modeling to interstellar polarization data (\citealt{2009ApJ...696....1D}; \citealt{2014ApJ...790....6H}). For the model with only silicate grains aligned, modeling requires $f_{\rm max}=1$, and for a mixture model of both carbon and silicate grains to be aligned, it requries $f_{\rm max}<1$ (\citealt{2009ApJ...696....1D}; \citealt{2018A&A...610A..16G}).

\section{Application to $\rho$ Oph-A} \label{sec:compare}
\subsection{Numerical setup}
\begin{figure}
    \centering
    \includegraphics[width=0.45\textwidth]{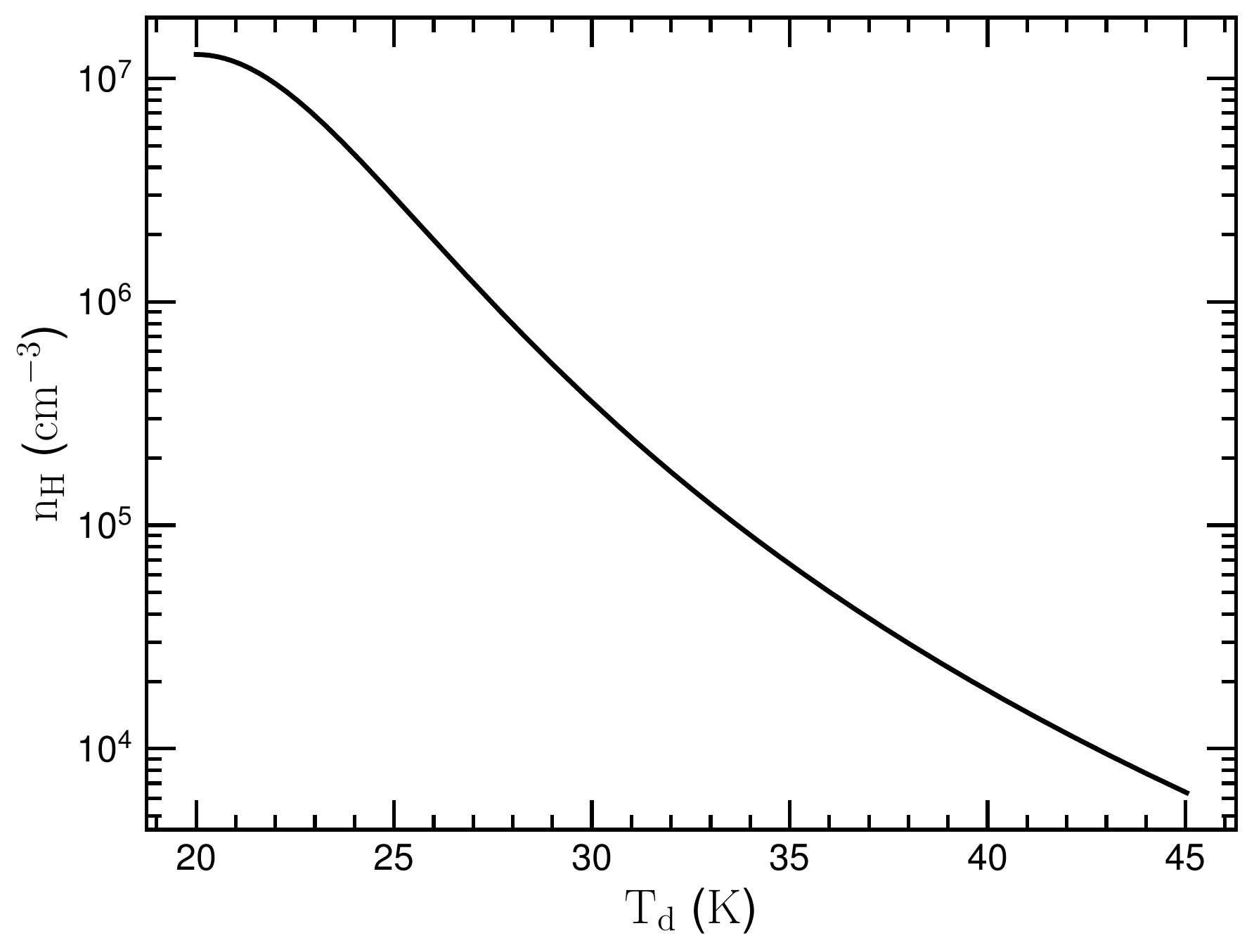}
    \caption{Relation between the local gas number density and the local dust temperature.}
    \label{fig:nH_Td}
\end{figure}

As discussed in Section \ref{sec:model}, the parameters of the model include the gas properties: gas number density ($n_{\H}$) and gas temperature ($T_{\rm gas}$); the dust properties: size ($a$), shape, internal structure (i.e., tensile strength $S_{\rm max}$) and size distribution power index ($\beta$); and the ambient properties: radiation field strength $U$ (which is in fact equivalent to the dust temperature $T_{\rm d}$), mean wavelength ($\bar{\lambda}$) and an anisotropy degree ($\gamma$) of the radiation field.   
Figure \ref{fig:polametric_maps}c shows that the gas is denser at the edge of the polarimetric map area and more diffuse close to the Oph S1 star. We derive the relation between the gas number density and the dust temperature by assuming that the dust temperature linearly decreases from $45\K$ down to $20\K$ at the edge of the polarimetric map area, and the gas number density is calculated from a spherical model as in Section 3.5 in \cite{2019ApJ...882..113S}. This relation is shown in Figure \ref{fig:nH_Td}. Throughout this work, we fix the value of the gas temperature as $T_{\rm gas}=20$ K, which is fairly common for dense molecular clouds.

In a dense molecular cloud, large grains are expected to be present thanks the coagulation process. We set the initial maximum value of grain size as $1\,\mu$m, then the RATD mechanism constrains the actual maximum value. The smallest value for the grain sizes is kept fixed at $10\,\AA$. The internal structure of grains is determined via their tensile strength (e.g., large composite grains have $S_{\rm max}\simeq 10^{7}\erg \cm^{-3}$, stronger grains have a higher value of $S_{\rm max}$), which is a free parameter. The grain-size distribution could change across the $\rho$ Oph-A cloud (\citealt{2015A&A...578A.131L}), thus we vary the power index $\beta$ as another free parameter. In our model, the local value of the radiation strength is determined by the dust temperature as shown in Figure \ref{fig:polametric_maps}d via the relation $T_{\rm d}=16.4 a^{1/15}_{-5}U^{1/6}$K (\citealt{2011piim.book.....D}). The mean wavelength is $\bar{\lambda}\simeq 0.3\mu$m corresponding to a B-like star with $T_{\ast}\simeq 1.5\times 10^{4}$K. The anisotropy degree is $\gamma= 1$ for the unidirectional radiation field from a nearby star. 
\begin{figure*}
    \centering
    \includegraphics[width=0.45\textwidth]{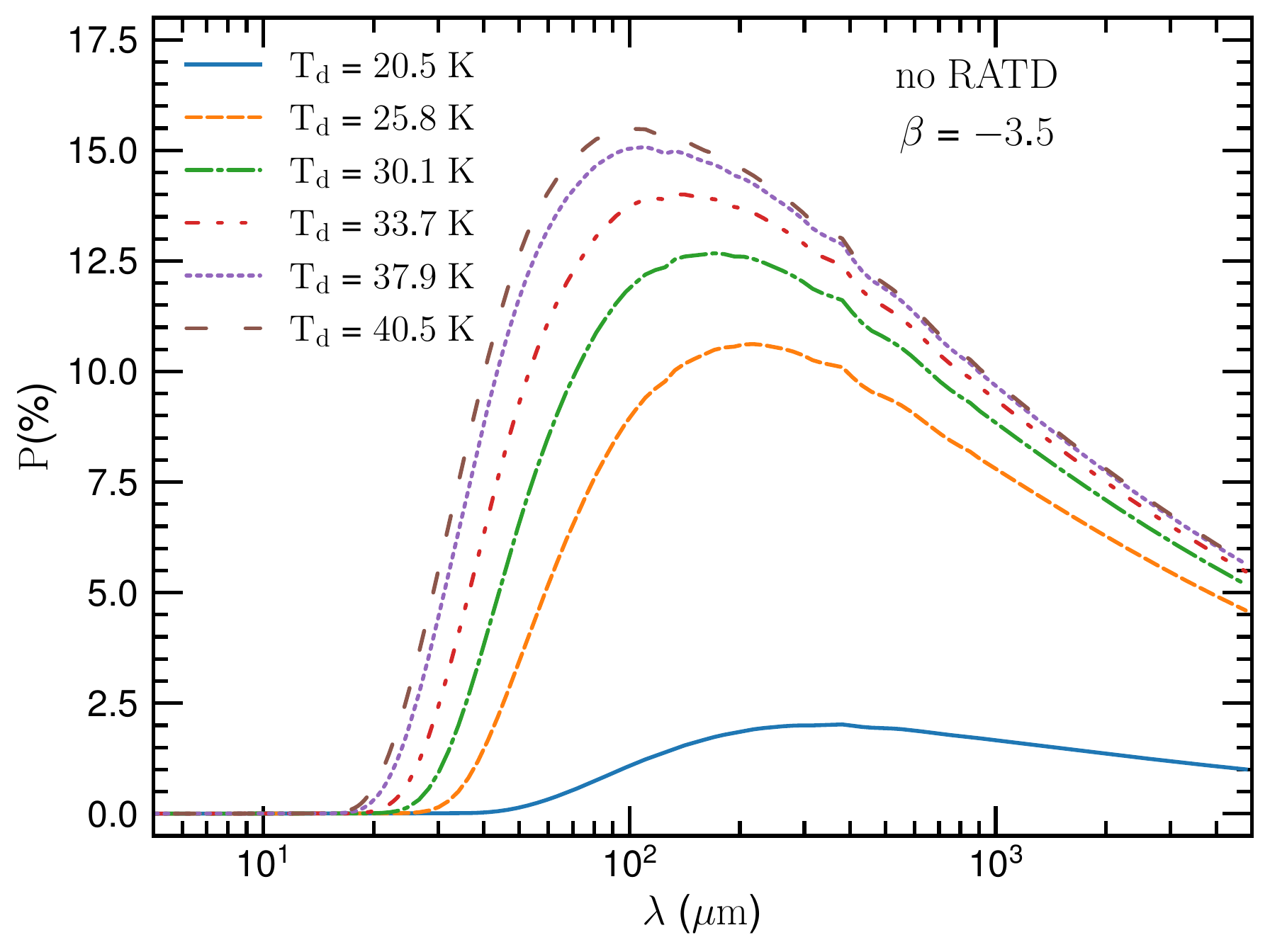}
    \includegraphics[width=0.45\textwidth]{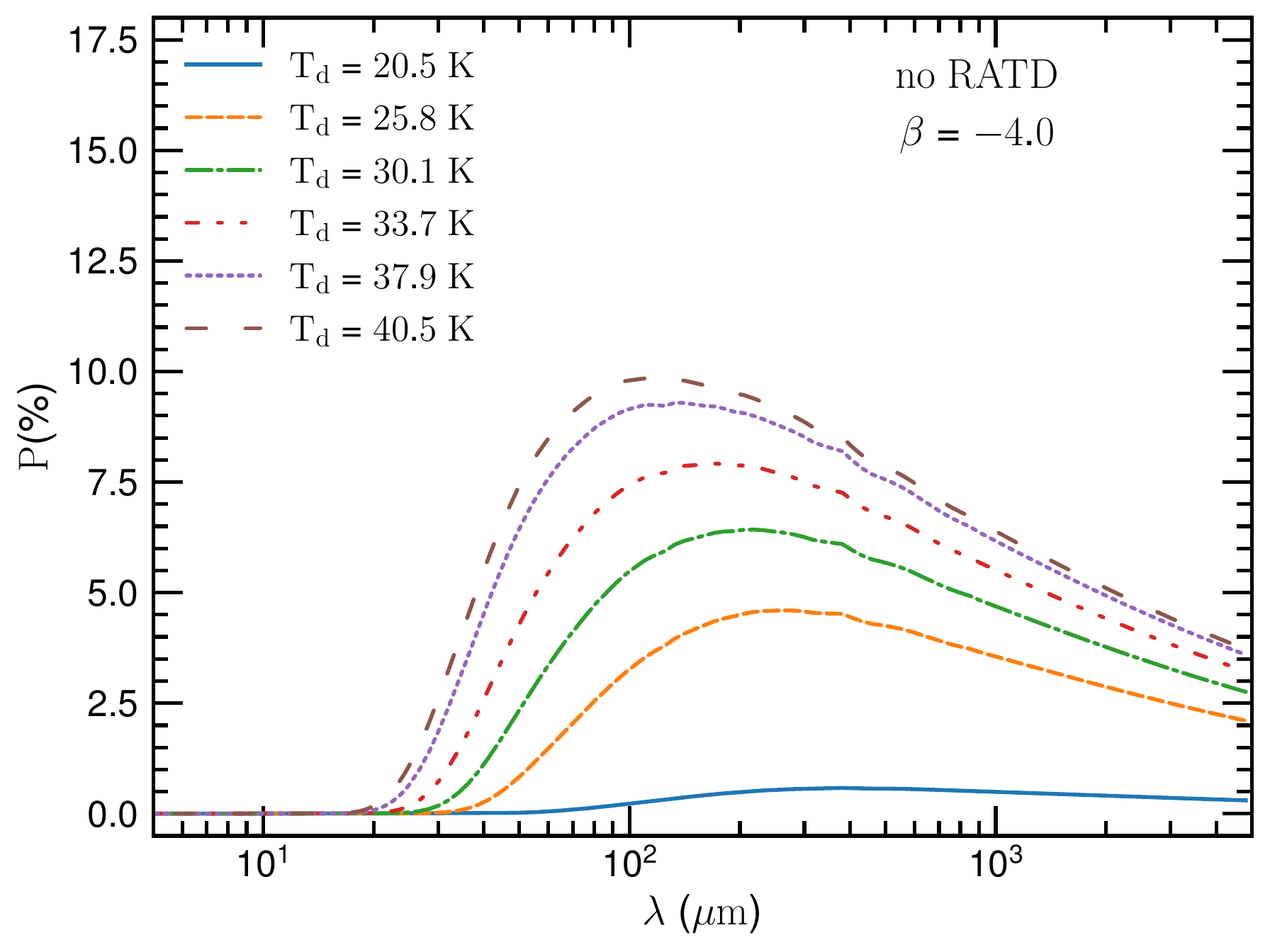}
    \caption{Polarization spectrum of thermal dust emission calculated from grain alignment by RATs only (without RATD) for different grain temperatures, $T_{\rm d}$, assuming the size distribution power index $\beta$=-3.5 (left panel) and $\beta$=-4.0 (right panel). Higher dust temperatures result in higher polarization degree and smaller peak wavelength of the spectrum. Steeper size distribution leads to lower polarization degree (right panel). Only silicate grains are assumed to be aligned, and carbonaceous grains are randomly oriented.}
    \label{fig:disruptionoff}
    \includegraphics[width=0.45\textwidth]{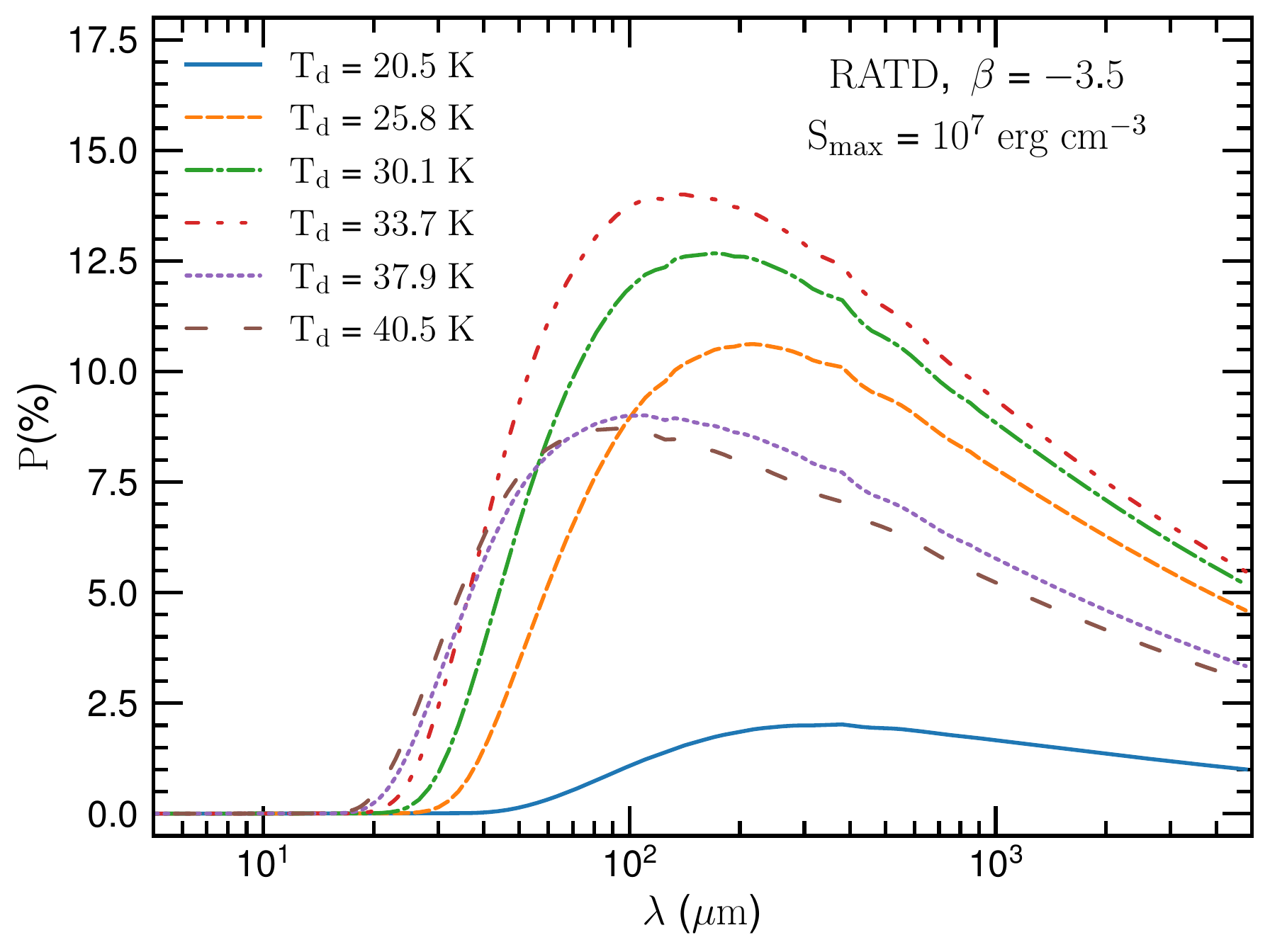}
    \includegraphics[width=0.45\textwidth]{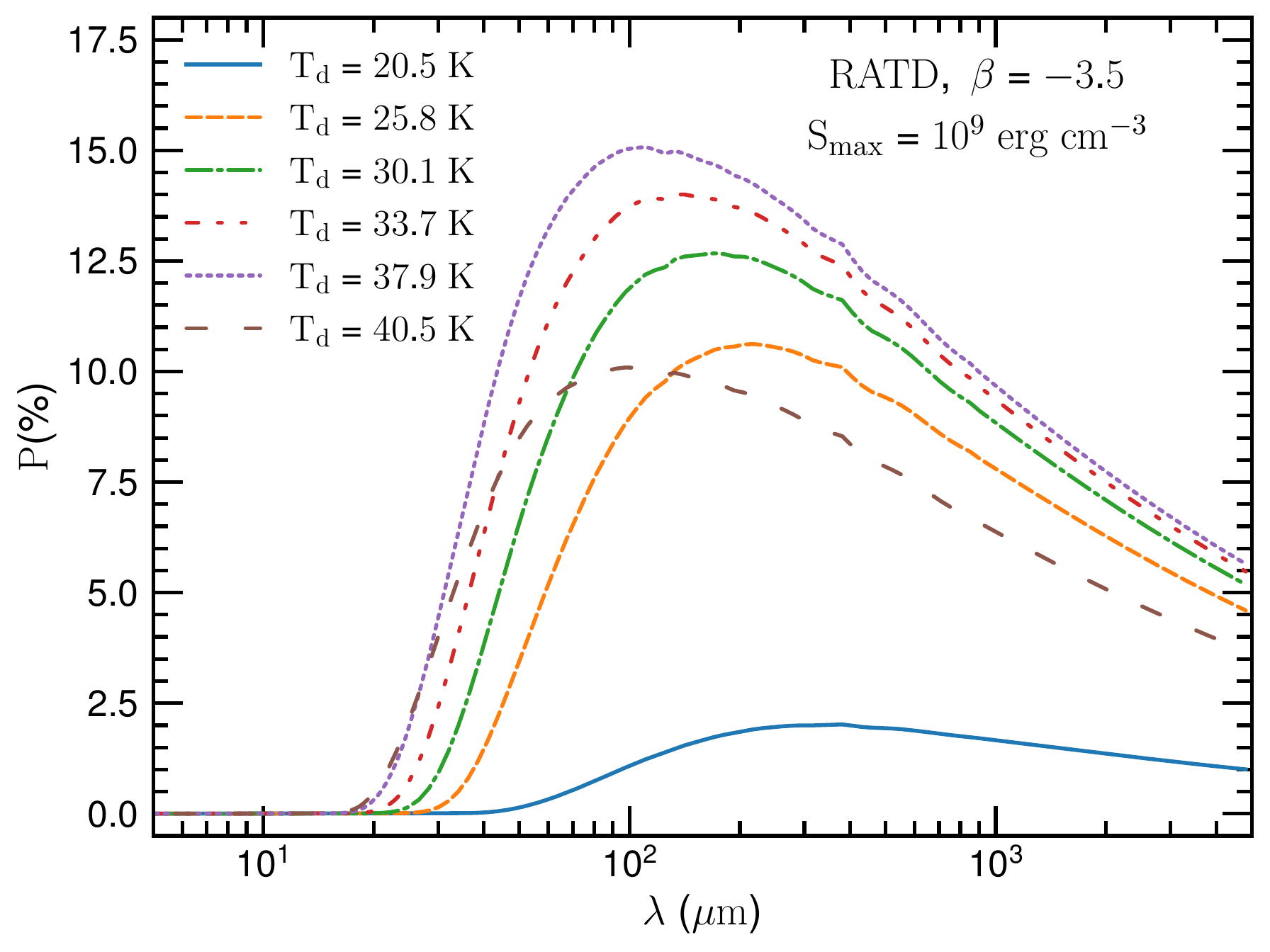}
    \caption{Polarization spectrum of thermal dust emission calculated with both grain alignment and disruption by RATs for two values of the tensile strength. The RATD effects decreases the polarization degree for $T_{\rm d}>33.7\ \K$ (left) and for $T_{\rm d}>37.9\K$ (right). The decline is more substantial for composite grains (left panel) than for more compact grain (right panel).}
    \label{fig:disruptionon}
    \includegraphics[width=0.46\textwidth]{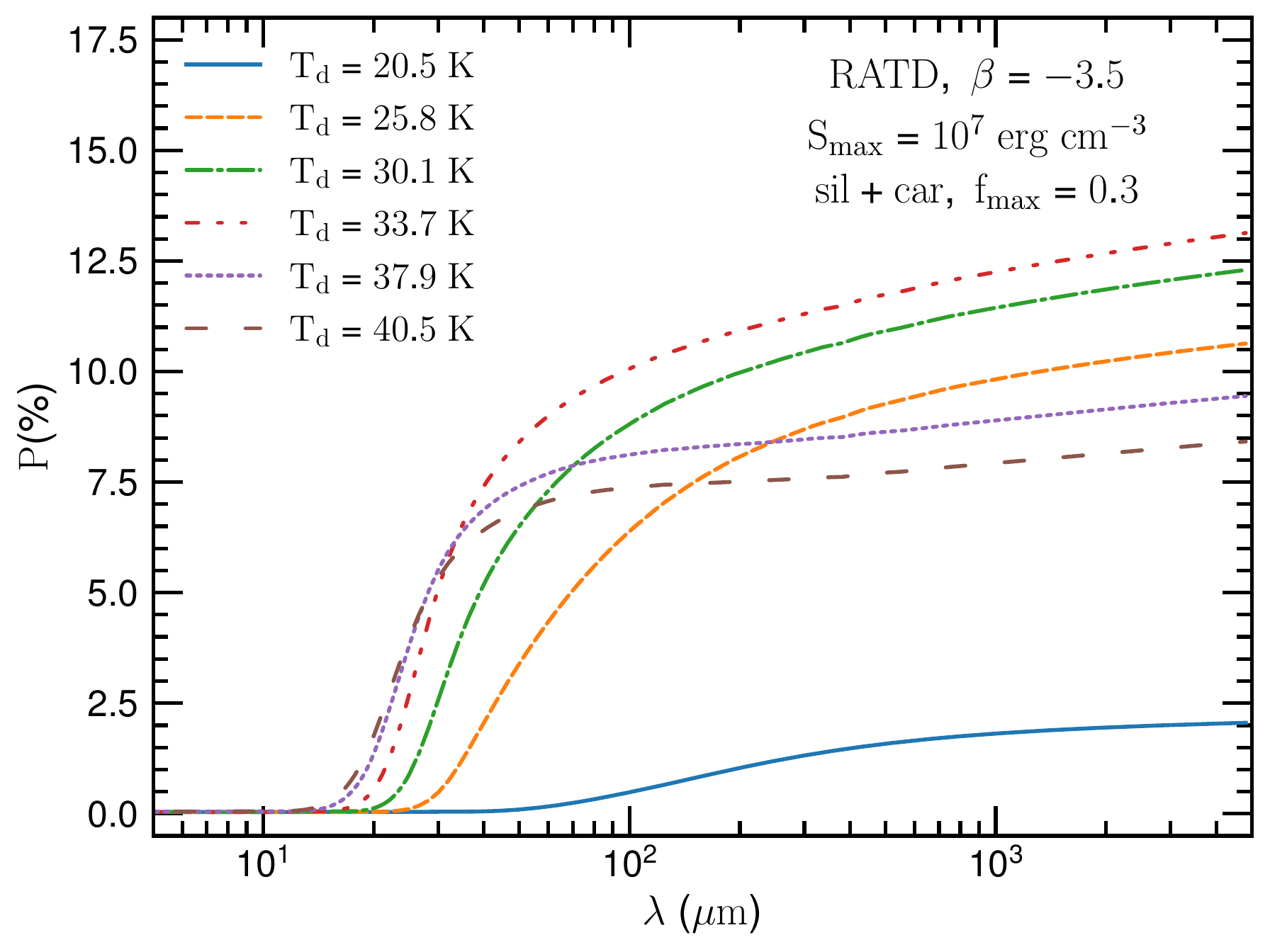}
    \includegraphics[width=0.45\textwidth]{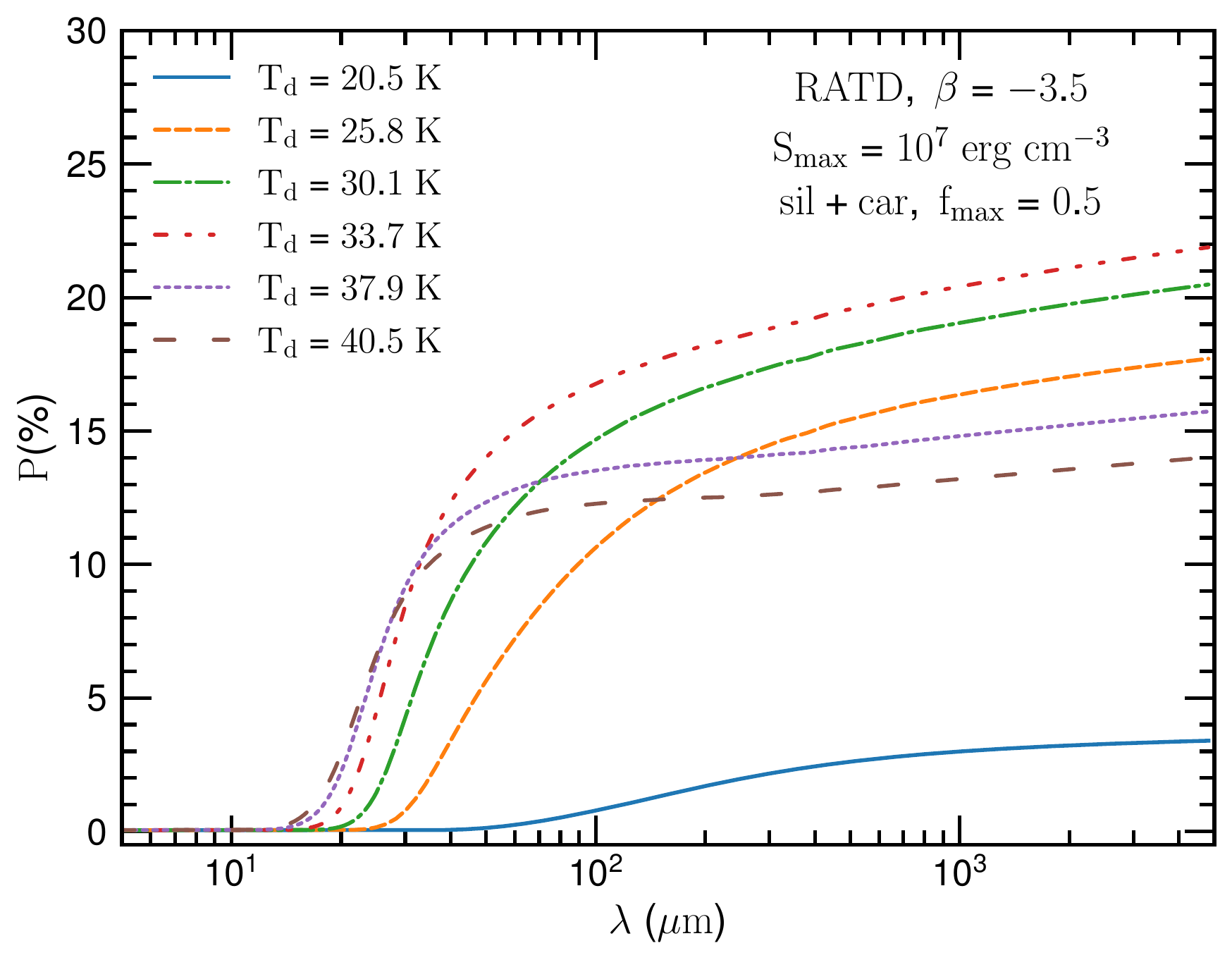}
    \caption{Same as Figure \ref{fig:disruptionon} (left panel) but for a mixture of silicate and carbon grains aligned with $f_{\rm max}=0.3$ (left panel) and $f_{\rm max}=0.5$ (right panel). The disruption effect also happens once $T_{\rm d}>34\K$. However, the shape shows a flat feature. Higher $f_{\rm max}$ leads to higher polarization degree.}
    \label{fig:disruptionon_silcar}
\end{figure*}

\begin{figure*}
    \centering
    \includegraphics[width=0.45\textwidth]{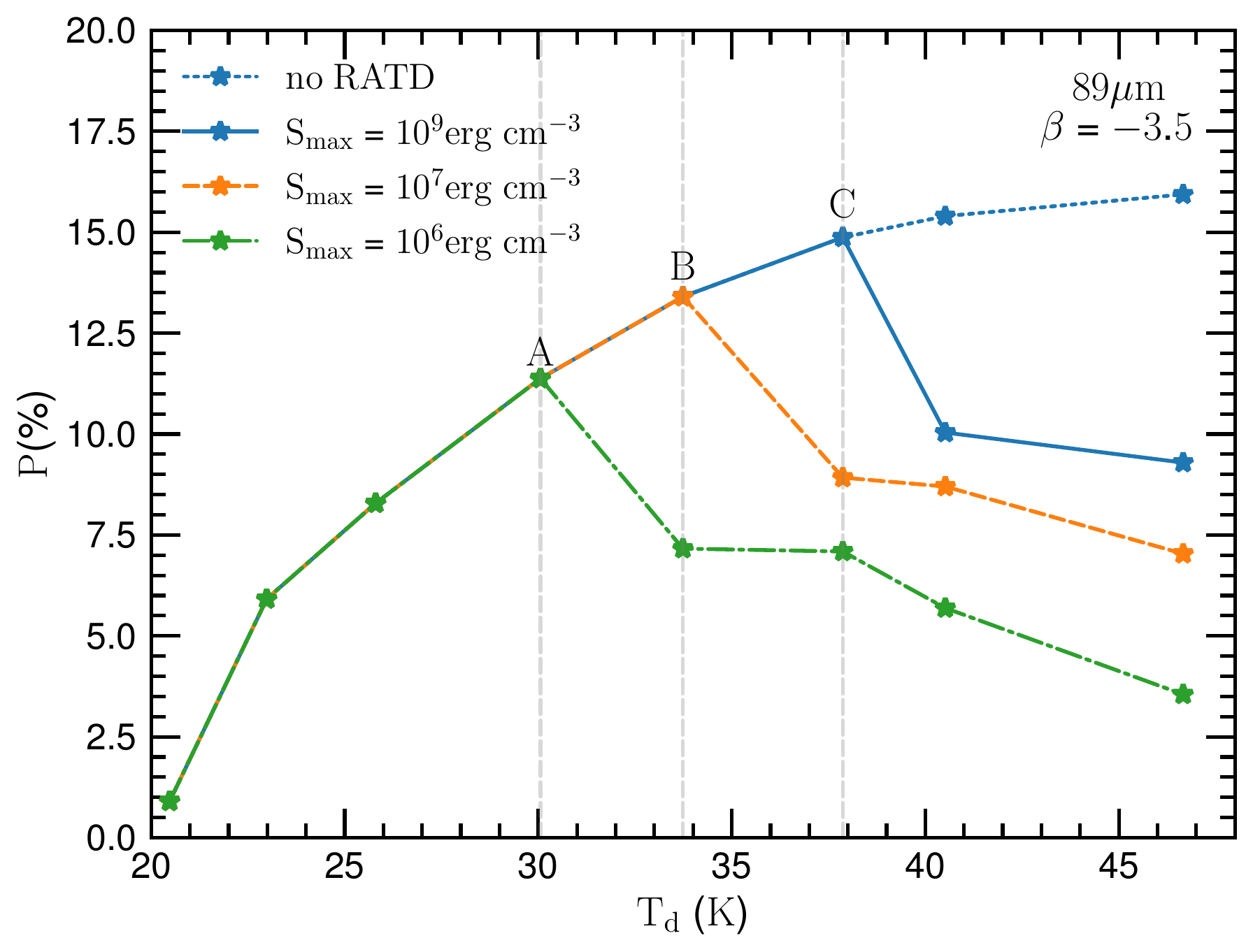}
    \includegraphics[width=0.45\textwidth]{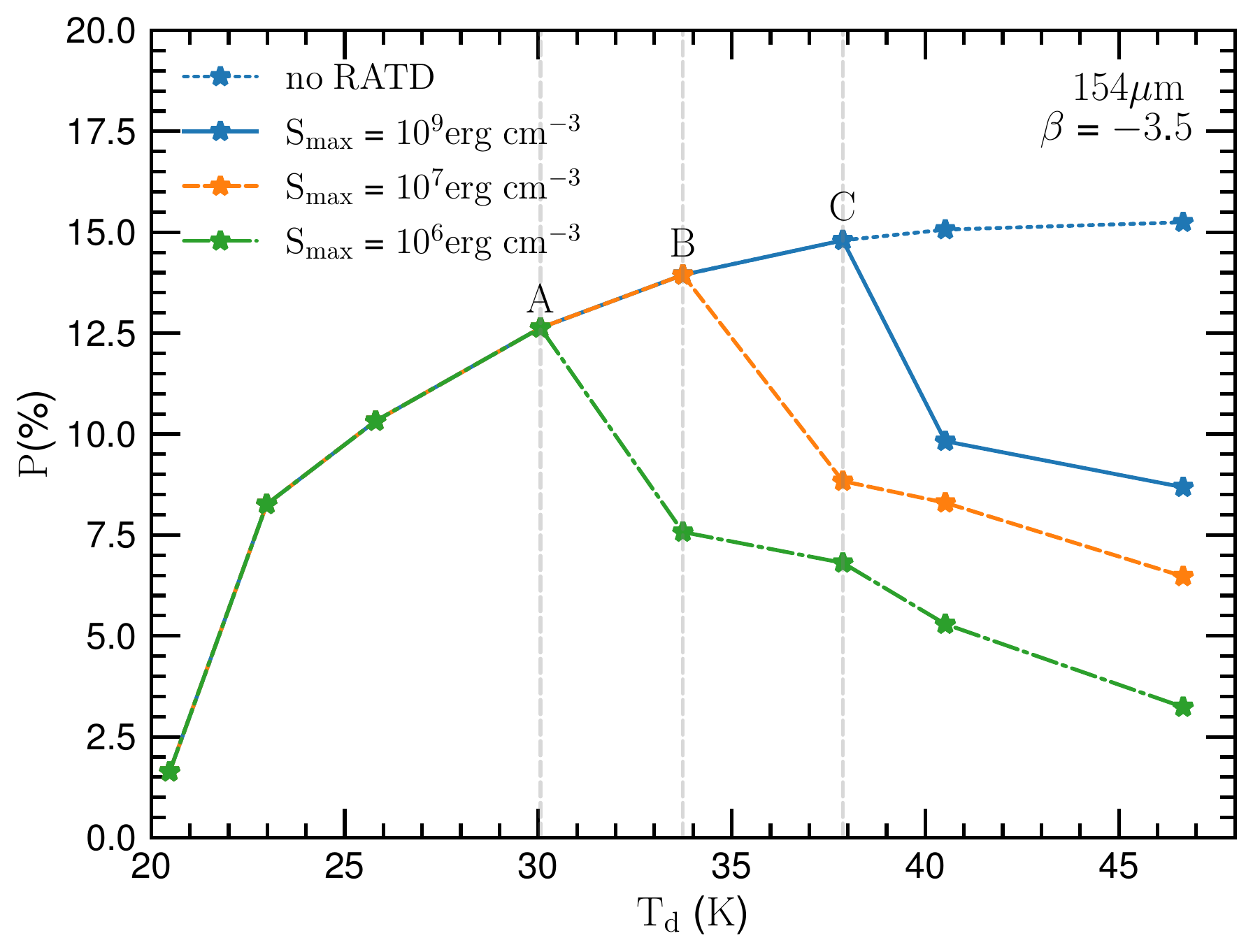}
    \caption{Variation of the polarization degree with the grain temperature, computed at 89 $\mu$m (left panel) and 154 $\mu$m (right panel) with and without RATD effect at a given grain-size distribution. Without RATD, the polarization degree monotonically increases as dust temperature increases (dotted blue line). With RATD, the polarization degree first increases and then decreases when the dust temperature exceeds some critical value. This value (labeled by A, B, and C), is lower for weak grains and larger for stronger grains. Only silicate grains are assumed to be aligned.}
    \label{fig:pol_Td_Smaxfixed}
    \includegraphics[width=0.45\textwidth]{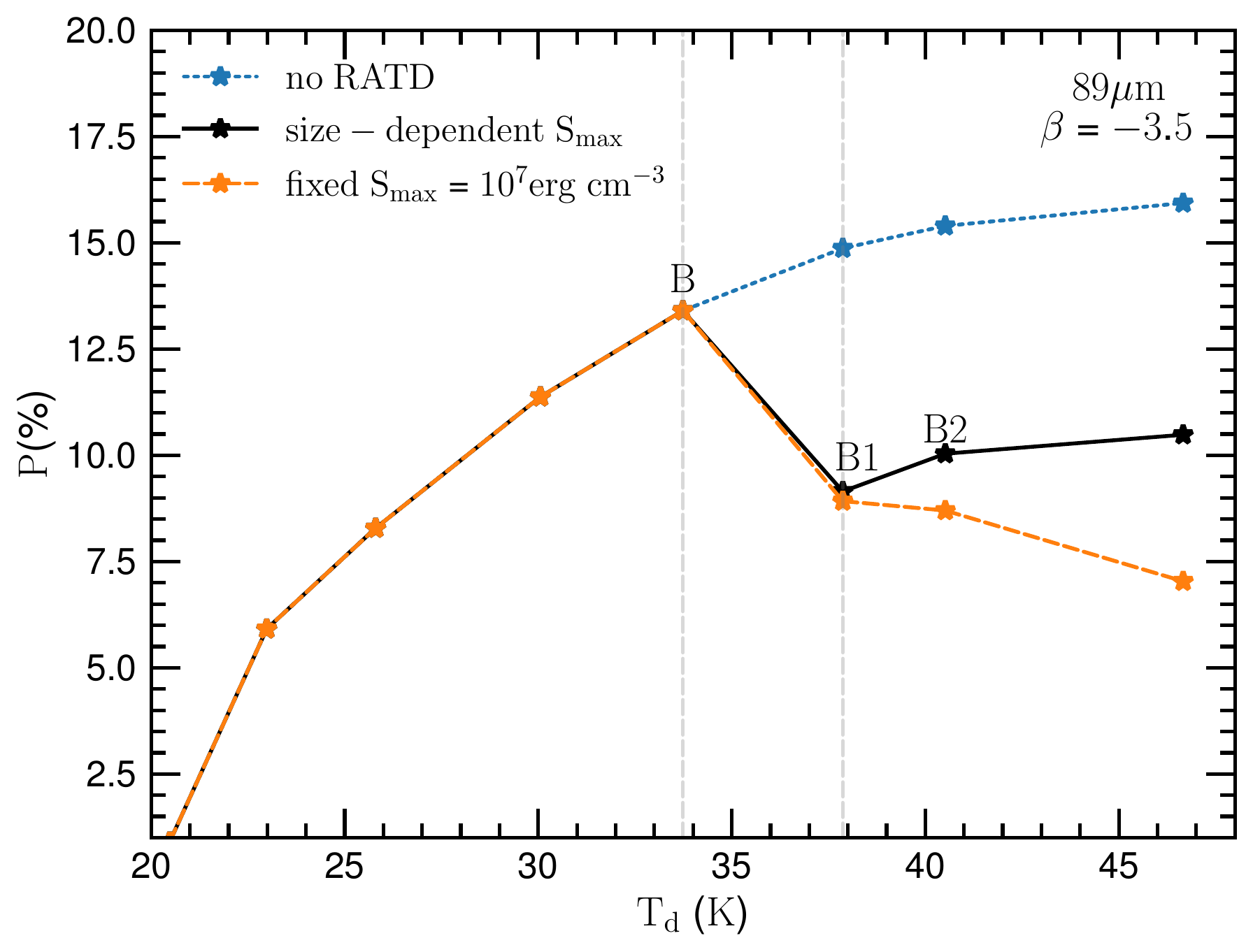}
    \includegraphics[width=0.45\textwidth]{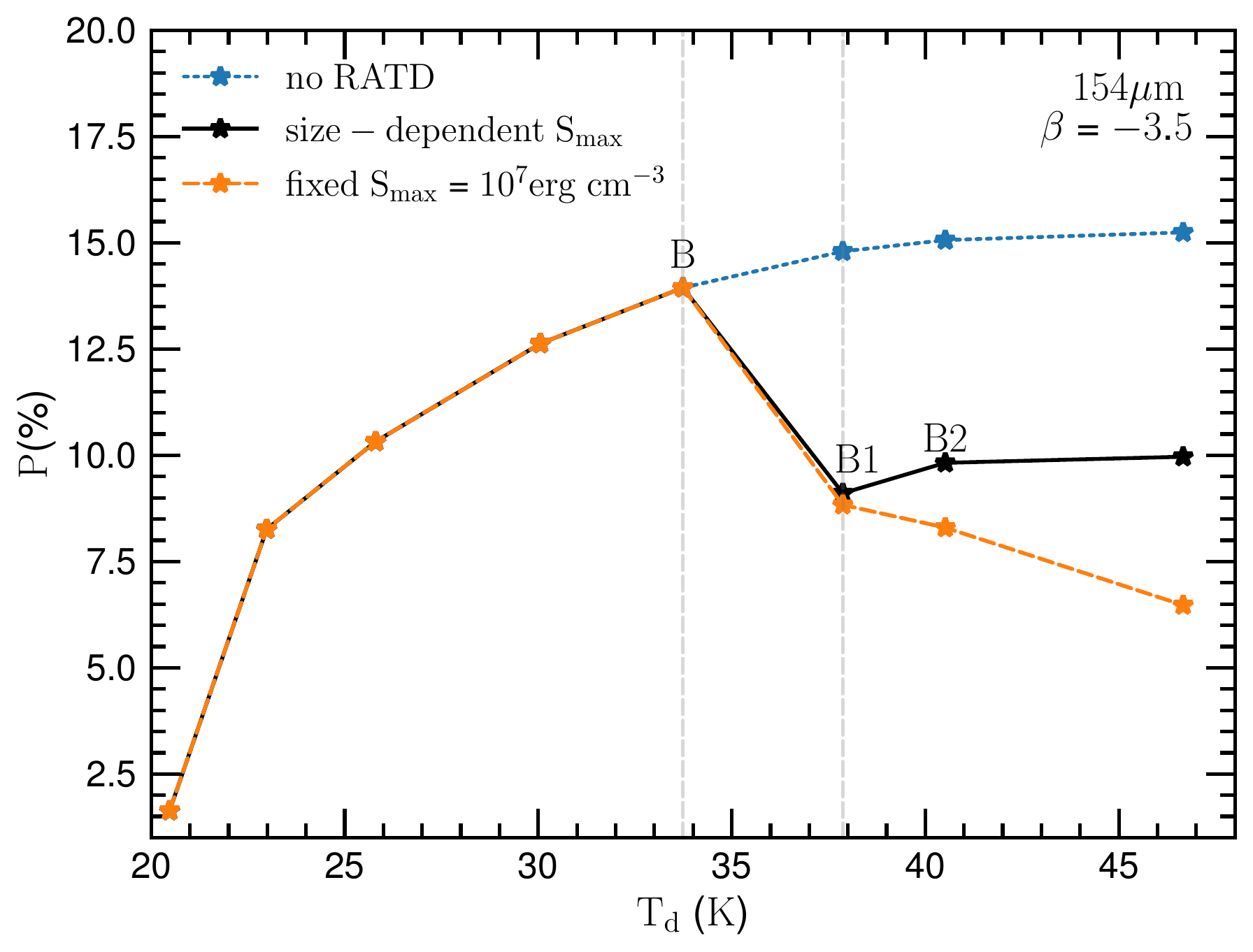}
    \caption{Effect of the size-dependent tensile strength on the fractional polarization emission of silicate grains. The dotted blue line and the dashed orange line are computed for a fixed $S_{\rm max}$ as in Figure \ref{fig:pol_Td_Smaxfixed}. The solid black line is the model prediction for a size-dependent $S_{\rm max}$ (see text for details).}
    \label{fig:pol_Td_Smaxvaried}
    \includegraphics[width=0.45\textwidth]{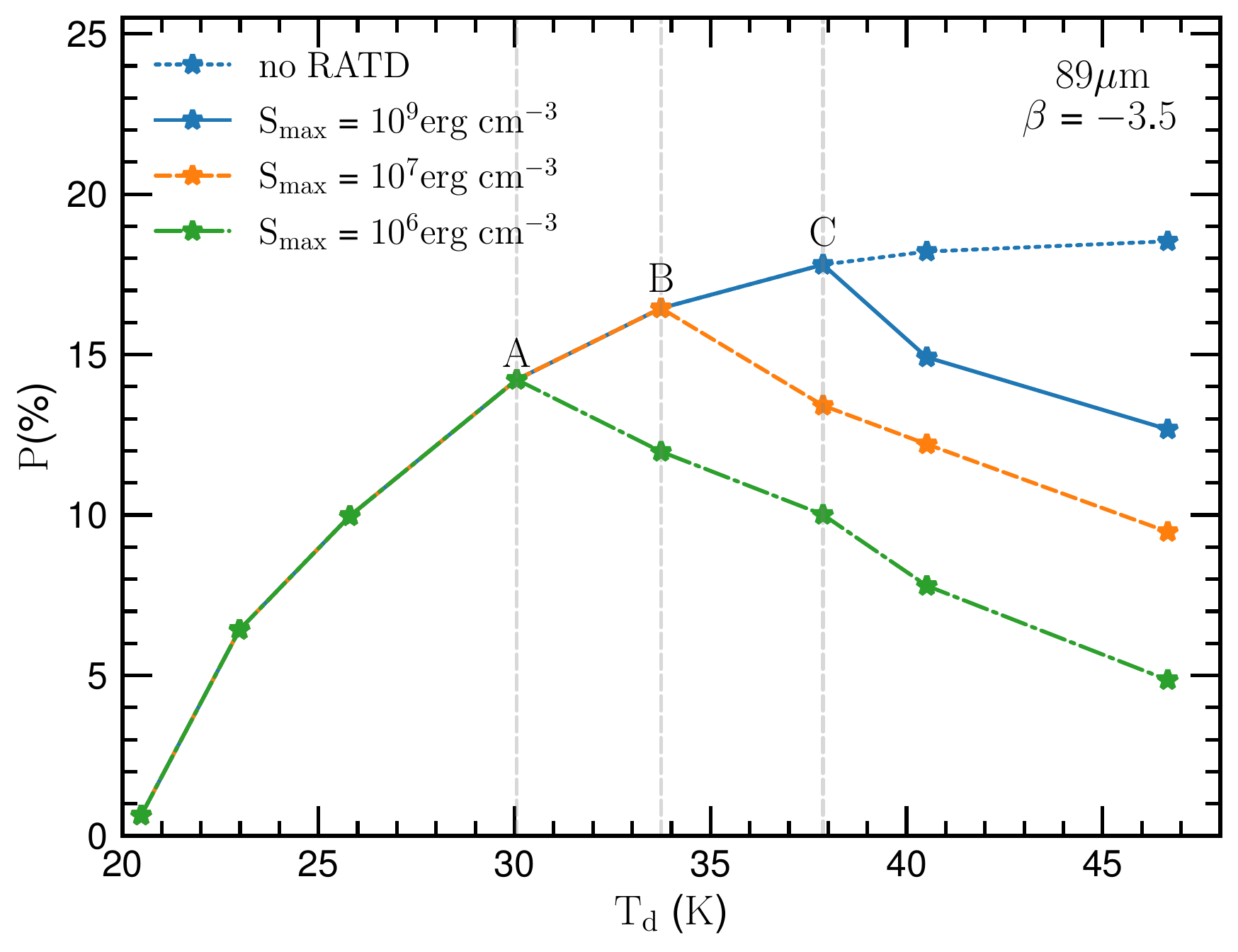}
    \includegraphics[width=0.45\textwidth]{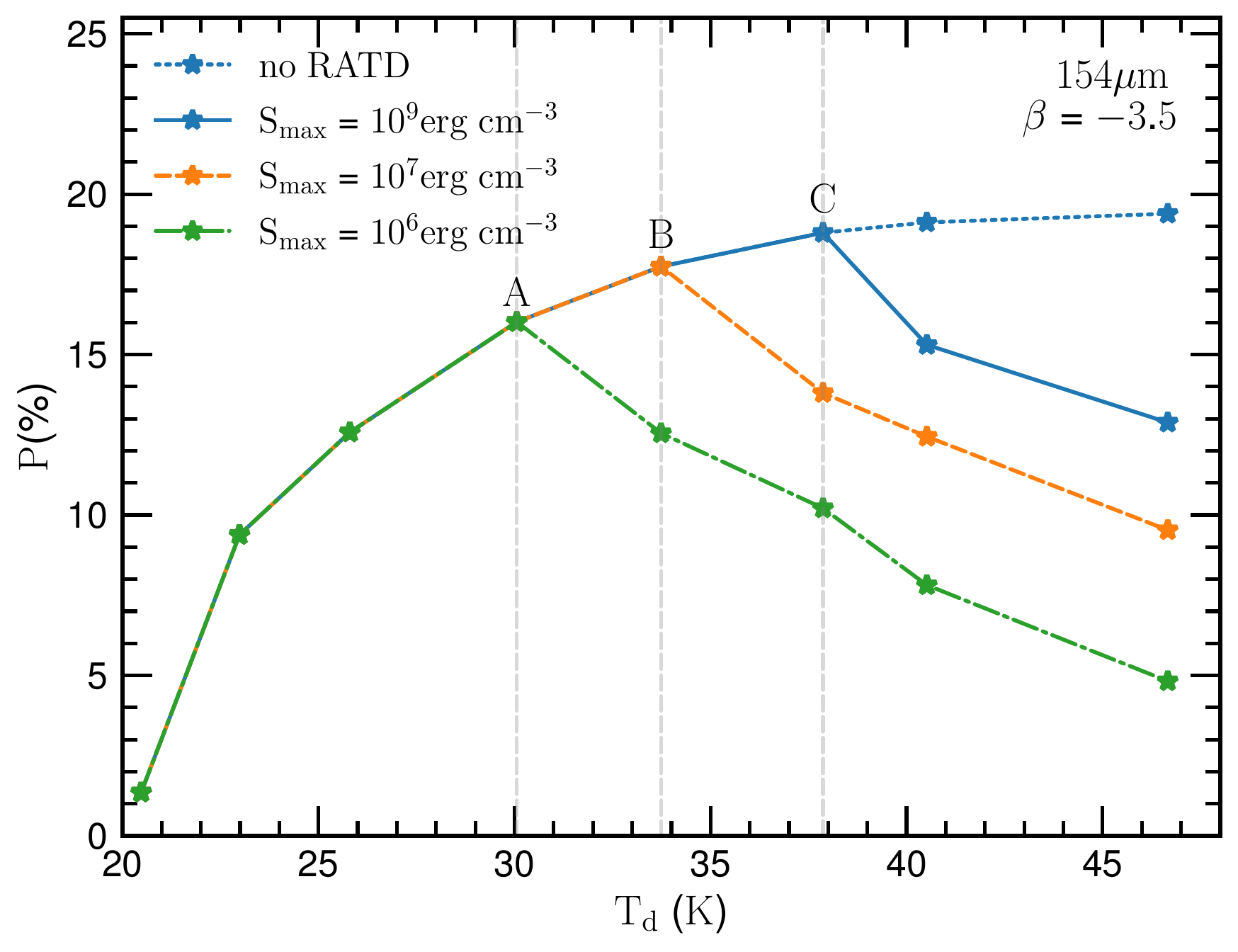}
    \caption{Same as Figure \ref{fig:pol_Td_Smaxfixed}, but both silicate and carbon grains are assumed to be aligned with $f_{\rm max}=0.5$. The trend and the critical temperature are the same but the decline is less steep and the polarization amplitude is higher than in the case of silicate grains alone.}
    \label{fig:pol_Td_Smaxfixed_carsil}
\end{figure*}

\subsection{Numerical results} \label{sec:numerical_results}
Here, we show the numerical results of the multi-wavelength polarization degree of thermal dust emission using RATs theory in two cases: without disruption (namely classical RATs) and with disruption for comparison. 

Figure \ref{fig:disruptionoff} shows the polarization spectra obtained with grain alignment by RATs only (without RATD), computed for several values of the dust temperature with different grain-size distributions, i.e., $\beta=-3.5$ (left panel) and $\beta=-4.0$ (right panel). One can see that (1) the polarization degree proportionally increases as the dust temperature increases, and (2) the polarization degree is lower for lower values of $\beta$ for the same $T_{\rm d}$. The first effect is due to the fact that a higher dust temperature (equivalent to higher radiation strength) causes larger torques acting on dust grains, which decreases the alignment size $a_{\rm align}$ and then increases the polarization degree of dust emission. Moreover, for a lower $\beta$, the dust mass contained in large grains is smaller, decreasing the polarization degree of the thermal dust emission that is dominantly produced by aligned, large grains. This explains the second effect.    

Figure \ref{fig:disruptionon} shows the polarization spectra obtained with both grain alignment and disruption by RATs (with RATD), assuming different values of tensile strength, i.e., $S_{\rm max}=10^{7}\erg \cm^{-3}$ (left panel) and $S_{\rm max}=10^{9}\erg \cm^{-3}$ (right panel). In the left panel, the low-$T_{\rm d}$ curves are the same with Figure \ref{fig:disruptionoff} (blue, orange, green, and dashed-dotted red lines). However, differing from Figure \ref{fig:disruptionoff}, the polarization degree decreases as dust temperature increases beyond a critical value (i.e., $\simeq 34\K$, the dotted violet and dashed brown lines). Higher $S_{\rm max}$ leads the disruption to occur at a higher critical dust temperature (i.e., $\simeq 38\K$, the dashed brown line). The reason is that dust grains, exposed to strong radiation (indicated by where the dust temperature is high, see Figure \ref{fig:polametric_maps}d), can be rotated extremely fast due to strong radiative torques while damping is inefficient (because of a low gas density, Figure \ref{fig:polametric_maps}c), resulting in radiative torques disruption (RATD) as described in Section \ref{sec:RATD}. For $T_{\rm d}$ lower than the critical temperature, on the contrary, the radiative torques are weaker, and the damping process is more substantial (because gas is denser) so that the RATD cannot occur and thus the results are the same as for the classical RATs calculations. 
The disruption leads to a drop in the polarization degree. The critical temperature above which RATD occurs and the level of the decline depend on the internal structure of the grains controlled by $S_{\rm max}$. The composite grains ($S_{\rm max}=10^{7}\erg \cm^{-3}$) are more easily disrupted, resulting in a significant decrease of the polarization degree (Figure \ref{fig:disruptionon}, left panel), than for the compact grains ($S_{\rm max}=10^{9}\erg \cm^{-3}$) (Figure \ref{fig:disruptionon}, right panel).      

Figure \ref{fig:disruptionon_silcar} shows the polarization spectrum for the case of mixed silicate and carbon grains in which both grain populations are aligned by RATs. Similar to Figure \ref{fig:disruptionon}, the disruption occurs for $T_{\rm d}>34\K$. In this case, the spectrum shows an increase and then a plateau feature, which differs from Figure \ref{fig:disruptionon}. The reason is that the polarization degree is the ratio of the polarized intensity ($I_{\rm pol}$) to the total intensity ($I_{\rm em}$) (Equation \ref{eq:pol_degree}). Since the $T_{\rm d}$ of silicate grains is lower than that of carbon grains, their spectrum slopes differ from each other. When only silicate grains are aligned the different spectrum slopes of $I_{\rm pol}$ to $I_{\rm em}$ result in a slope in polarization spectrum (see e.g., Figure \ref{fig:disruptionon}). When both silicate and carbon grains are aligned, $I_{\rm pol}$ and $I_{\rm em}$ differ by a factor of degree of grain alignment, which results in a flat spectrum. The degree of grain alignment is defined by $f_{\rm max}$ (Equation \ref{eq:fa}). For a combination of carbon and silicate grains, the non-perfect aligned grains ($f_{\rm max}<1$) can reproduce observation (see e.g., \citealt{2009ApJ...696....1D}; \citealt{2018A&A...610A..16G}). Grains with higher value of $f_{\rm max}$ (right panel) produce more polarized thermal emission than for a lower value of $f_{\rm max}$ (left panel).  

Figure \ref{fig:pol_Td_Smaxfixed} shows the polarization degree at 89 $\mu$m (left panel) and 154 $\mu$m (right panel) with respect to dust temperature. In the case without RATD (dotted lines), the polarization degree first increases rapidly with increasing dust temperature and then slowly changes (as shown in Figure \ref{fig:disruptionoff}). Accounting for RATD, the polarization degree first increases with $T_{\rm d}$ and then rapidly declines once the dust temperature exceeds a critical value, which depends on the grains' tensile strength as shown in Figure \ref{fig:disruptionon}. The critical dust temperature is lower for weaker grains (i.e., lower value of $S_{\rm max}$) because of an effective disruption, which leads to a deeper decrease of the polarization degree in comparison to stronger grains.

Above, we assume that all grains have the same tensile strength ($S_{\rm max}$). However, the tensile strength of composite grains scales with the radius of monomoers as $a_{p}^{-2}$ (see \citealt{2019ApJ...876...13H} for a detailed demonstration), which implies that large grains (comprising many monomers) are more breakable than smaller grains presumably having a compact structure. As an example, we set $S_{\rm max}=10^{7}\erg \cm^{-3}$ for all grains with size $a\geq 0.1\,\mu$m while $S_{\rm max}=10^{9}\erg \cm^{-3}$ for smaller grains. The results are shown in Figure \ref{fig:pol_Td_Smaxvaried}, black solid lines. The trend of the polarization degree also shows an increase and decrease feature to dust temperature. However, for $T_{\rm d}>T_{\rm crit}$ (i.e., denoted by the position B), its amplitude is higher than in the case of fixed $S_{\rm max}$. The reason is as follows. When the RATD does not occur, the polarization is higher for higher dust temperature/stronger radiation. When the dust temperature is just enough for RATD to occur, the disruption mostly effects the largest grains (i.e., low $S_{\rm max}=10^{7}\erg \cm^{-3}$ in this example), so that the curve follows the case of fixed $S_{\rm max}=10^{7}\erg \cm^{3}$ (e.g., BB1 slope). As the dust temperature increases, the RATD can affect smaller grains (i.e., higher $S_{\rm max}$). Because the decline of polarization is smaller for higher $S_{\rm max}$, there is a short increasing interval in the polarization (see the B1, B2 segment). Finally, once RATD only effects "strong" grains, the trend of the polarization follows the fixed $S_{\rm max}=10^{9}\erg \cm^{-3}$ case, as shown in Figure \ref{fig:pol_Td_Smaxfixed}.

Figure \ref{fig:pol_Td_Smaxfixed_carsil} shows the polarization degree of thermal dust as a function of the grain temperature, assuming that both carbon and silicate grains are aligned. The results generally show that the polarization degree drops at the same critical dust temperature as Figure \ref{fig:pol_Td_Smaxfixed} in which only silicate grains are aligned. However, the mixed grain model results in higher polarization, as well as in less decline than in the case of single grains due to the contribution of aligned carbon grains. With the $T_{\rm d}-n_{\H}$ relation, we varied the value of $n_{\H}$ by 10$\%$ but we do not see a significant change (i.e., the correlation coefficient is $\simeq 0.99$). However, another relation of $n_{\H}$ to $T_{\rm d}$ could affect more significantly.          

\subsection{Interpretation of observations}
 \begin{figure*}
    \centering
    \includegraphics[width=0.9\textwidth]{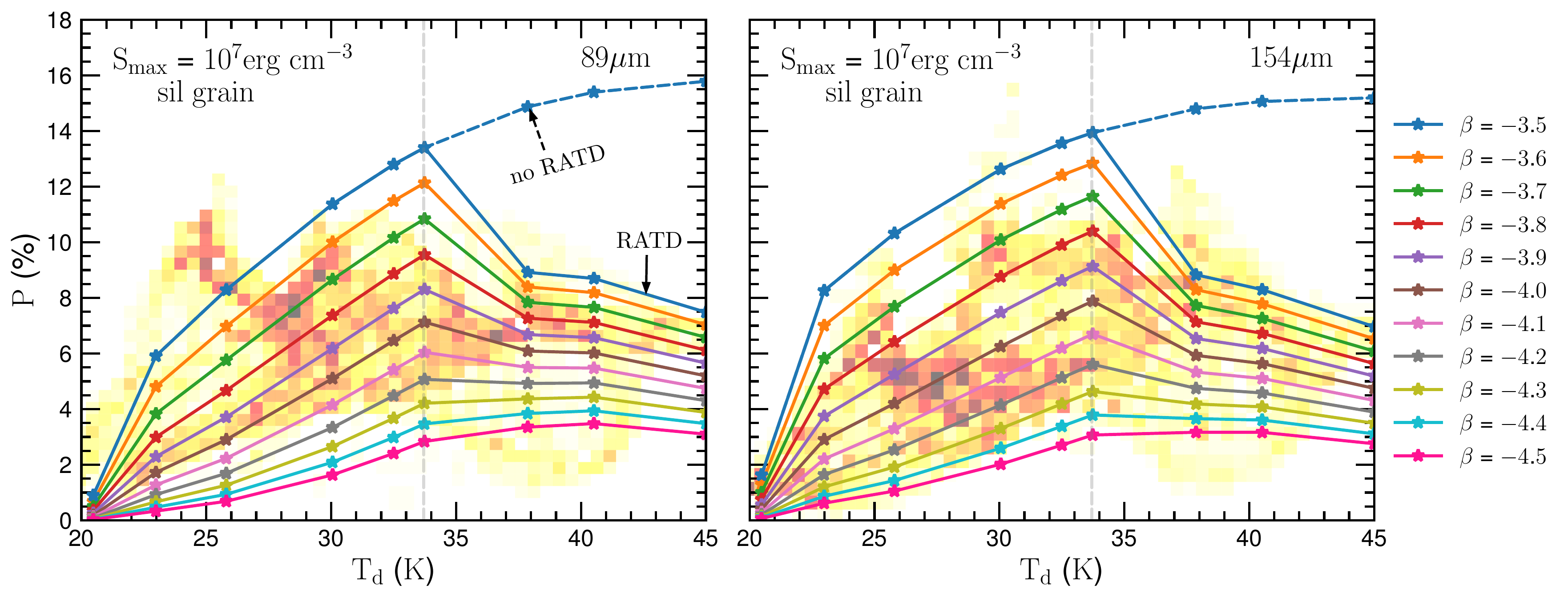}
    \caption{Comparison of the polarization degree of dust emission from our models with observations at 89 $\mu$m (left panel) and at 154 $\mu$m (right panel). Background colored points are the observational polarization degree (Figure \ref{fig:hist2d_maps}). Colored lines are the models for different value of the power-index $\beta$. The dashed line shows the results without RATD, while solid lines show results with RATD. Tensile strength is $S_{\rm max}=10^{7}\erg \cm^{-3}$. Only silicate grains are aligned, and carbons are randomly oriented.}
    \label{fig:fits_obs_Smax1e7}
\end{figure*}

\begin{figure*}
    \centering
    \includegraphics[width=0.9\textwidth]{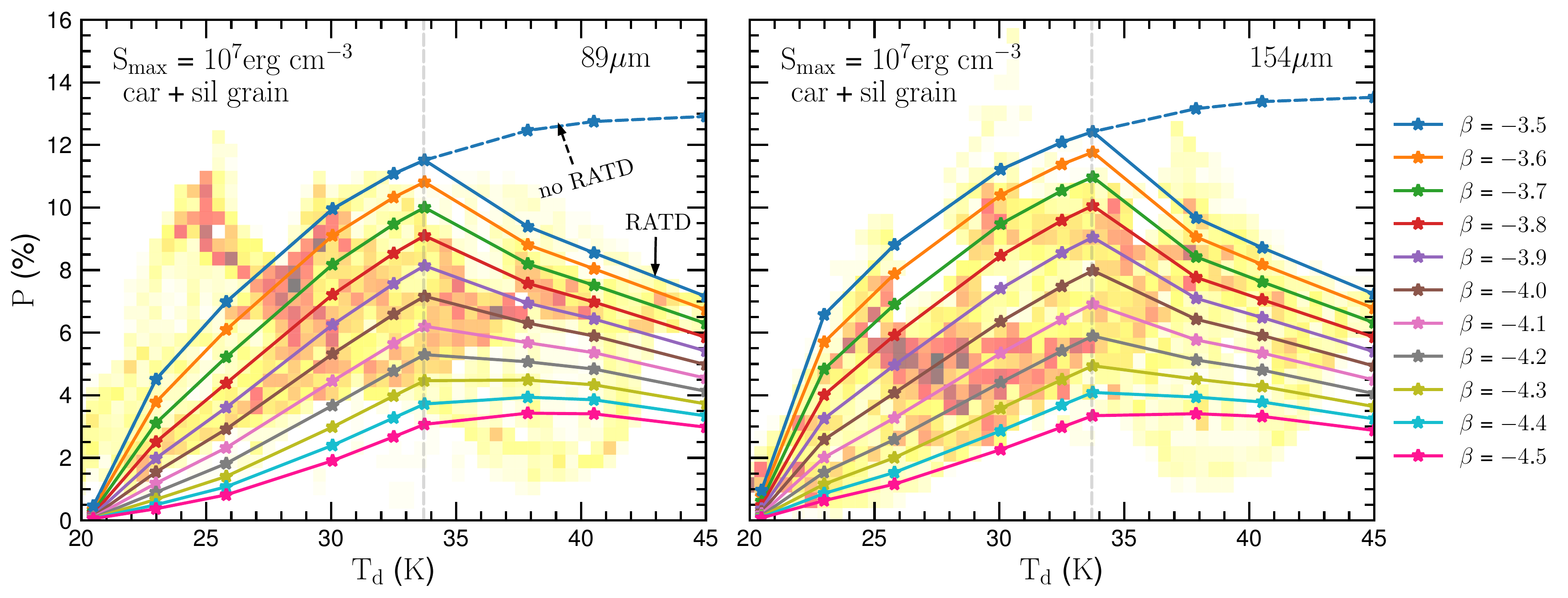}
    \caption{Similar to Figure \ref{fig:fits_obs_Smax1e7} but for a combination of aligned carbonaceous and silicate grains. The same range of $\beta$ can match better to the observations with the degree of alignment $f_{\rm max}=0.35$.}
    \label{fig:fits_obs_Smax1e7_carsil}
\end{figure*}
Since the $T_{\rm crit}\simeq 30-34\K$ critical dust temperature above which the polarized thermal dust emission drops for $S_{\rm max}=10^{6}-10^{7}\erg \cm^{-3}$ (see Figures \ref{fig:pol_Td_Smaxfixed}, \ref{fig:pol_Td_Smaxfixed_carsil}) is consistent with observations (see Figure \ref{fig:hist2d_maps}), Figure \ref{fig:fits_obs_Smax1e7} shows only the numerical results for $S_{\rm max}=10^{7}\erg \cm^{-3}$ with a variation of the silicate grain-size distribution power-index $\beta$ overlaid over the observational data. For illustration, we also show the results from the RATs model without the disruption effect (dashed line), which we denote as classical RATs theory. Since RATs theory implies that stronger radiation strength causes more torques acting on grains, which results in higher polarization, then the classical RATs model can only lead to an increase of the dust polarization degree with respect to dust temperature and fails to explain its decrease beyond $T_{\rm crit}$ dust temperature (as discussed in Section \ref{sec:numerical_results}). When the rotational disruption mechanism is incorporated into RATs (solid lines), the model can reproduce the increase and decrease features from observations. For $T_{\rm d}<T_{\rm crit}$, the disruption does not proceed; hence, the model is exactly as classical RATs, which accounts for the increase of the polarization degree. For $T_{\rm d}\geq T_{\rm crit}$, on the contrary, the disruption occurs so that large grains are disrupted into many smaller fragments. The enhancement of smaller grains causes a decrease in the polarization degree at these FIR wavelengths.

Furthermore, different solid lines correspond to different values of the grain-size distribution power-index $\beta$. In the case of silicate grains alone, a simple $\chi^{2}$ calculation as in Table \ref{tab:chi2} shows that it gets minimum at $\beta \simeq -4.0$ to observational data at 89 $\mu$m, while it is slightly lower as $\beta \simeq  -4.1$ to 154 $\mu$m data. Hence, the slope of the size distribution is steeper than the MRN size distribution for the standard interstellar medium (\citealt{1977ApJ...217..425M}), which is evidence of enhancement of small grains by RATD. 154 $\mu$m probes different (more embedded) layers of the $\rho$ Oph-A than the 89 $\mu$m observations do, thereby the size-distribution could be slightly different. Polarimetric data at longer wavelengths (e.g., 850 $\mu$m JCMT/SCUBA-2 observations, see \citealt{2018ApJ...859....4K}), which trace bigger grain size, are desired to get a more comprehensive picture. 

Figure \ref{fig:fits_obs_Smax1e7_carsil} shows the comparison for a mixture of carbon and silicate grains with respect to observations. As shown in Section \ref{sec:numerical_results}, both grain-size distribution ($\beta$) and degree of alignment ($f_{\rm max}$) control the amplitude of the polarization degree, but they do not affect the spectrum trend. We found that the same range of $\beta$ as in Figure \ref{fig:fits_obs_Smax1e7} also nicely fits the observational trend with $f_{\rm max}\simeq 0.35$. In this case, the $\chi^{2}$ calculation in Table \ref{tab:chi2} indicates that $\beta \simeq -4.0$ and $\beta \simeq -4.1$ also result in minimum $\chi^{2}$ to observed data at $89\,\mu$m and $154\,\mu$m, respectively.

\begin{table}
\centering
\caption{$\chi^{2}$ of the models with a single aligned silicate grains (Figure \ref{fig:fits_obs_Smax1e7}) and a combination of aligned carbonaceous and silicate grains (Figure \ref{fig:fits_obs_Smax1e7_carsil}) to observations computed by}
\label{tab:chi2}
\begin{tabular}{ccc|cc}
\multicolumn{5}{c}{$\chi^{2}=\frac{1}{N}\sum^{N}_{i} (P^{i}_{\rm obs} - P_{\rm mod})^{2}/P^{i}_{\rm obs}$} \\
\multicolumn{5}{c}{with $N$ the number of data points} \\
\\
\hline
{}&\multicolumn{2}{c|}{$\chi^{2}$ (89$\,\mu$m, $f_{\rm max}=1$)} & \multicolumn{2}{c}{$\chi^{2}$ (154$\,\mu$m, $f_{\rm max}=0.35$)} \\
$\beta$ & sil grain & car+sil grain & sil grain & car+sil grain \\
\hline
-3.5 & 4.99 & 3.41 & 10.62 & 7.40 \\ 
-3.6 & 3.45 & 2.65 & 7.60 & 5.69 \\
-3.7 & 2.36 & 2.03 & 5.12 & 4.15 \\
-3.8 & 1.68 & 1.59 & 3.25 & 2.86 \\
-3.9 & 1.37 & 1.35 & 1.98 & 1.89 \\
-4.0 & 1.34 & 1.31 & 1.26 & 1.28 \\
-4.1 & 1.53 & 1.44 & 1.01 & 1.02 \\
-4.2 & 1.86 & 1.71 & 1.09 & 1.04 \\
-4.3 & 2.26 & 2.08 & 1.41 & 1.28 \\
-4.4 & 2.71 & 2.50 & 1.84 & 1.67 \\
-4.5 & 3.15 & 2.94 & 2.33 & 2.12 \\
\hline
\end{tabular}
\end{table}

\subsection{Limitations of the model}
Our model's primary and  most sensitive input parameters are the local gas column density and the local dust temperature. The first controls the damping process of the rotating grains, while the second defines the angular rotational rate of grains. The value for the gas column density is derived from a spherical model, whereas the value for the dust temperature is adopted from observations. Therefore, our results contain uncertainties, and we would like to address here the main limitations of our model. First, the adopted value of dust temperature is, in fact, the projection on the plane of the sky, the actual value could be higher than these. Second, the dust temperature and gas density maps are derived from only three FIR bands of \textit{Herschel}/PACS ($60\,\mu$m, $100\,\mu$m, and $160\,\mu$m). The derivation could be more accurate if the (sub)millimeter and radio bands are taken into account as it was in \cite{2019ApJ...872..187C}. However, we expect that accounting for local variations of dust temperature and gas number density could explain the observational scatter, but should not change the trend or our conclusions. 

Because our main input parameters are the local values, our prescription will be easy to incorporate into more elaborate models that have better physical treatments for the gas and dust properties, such as 3D radiative dust modeling codes (e.g., \citealt{2012ascl.soft02015D}; \citealt{2015A&A...578A.131L}). 

Finally, we note that the magnetic field geometry is assumed to not vary along the line of sights toward $\rho$ Oph-A in the modeling. The effect of turbulent magnetic field would reduce the polarization degree predicted, but the trend $P(\%)$ vs. $T_{\rm d}$ is not affected. Nevertheless, the inferred magnetic field direction shown in Figure 2 in \cite{2019ApJ...882..113S} indicates the coherent magnetic stream lines in $\rho$ Oph-A. The turbulence, therefore, may occur at very small scale.

\section{Summary and conclusions} \label{sec:discussion}
We showed and interpreted the relation between the fractional polarization of thermal dust emission and dust temperature in $\rho$ Oph-A molecular cloud using the archival SOFIA/HAWC+ observations at 89 $\mu$m and 154 $\mu$m. The observed fractional polarization first increases with increasing dust temperature and then decreases once the dust temperature exceeds $\simeq 25-32\K$. This is similar to what seen in {\it Planck} data for other clouds (\citealt{2018arXiv180706212P}). This trend differs from the prediction by the classical RAT theory and represents a challenge to grain alignment theory. 

We calculated the polarization degree of thermal dust emission by simultaneously considering grain alignment and rotational disruption (RATD) induced by RATs. The RATD mechanism relies on the extremely fast rotation of large grains exposed in a strong radiation field (or high dust temperature in equivalent). For sufficiently high rotation rate, the centrifugal force can exceed the binding force that holds the grain's structure and disrupts the large grain into smaller species. Since RATs are stronger for larger grains, the RATD mechanism constrains the upper limit for the grain size distribution. The efficiency of RATD also depends on the grain tensile strength ($S_{\rm max}$), which is determined by its internal structure. {A compact structure grain has a high value of $S_{\rm max}\simeq 10^{9}\erg \cm^{-3}$, while a composite structure has a lower value of $S_{\rm max} \simeq 10^{6}-10^{7}\erg \cm^{-3}$, and a porous structure has even lower $S_{\rm max}<10^{6}\erg \cm^{-3}$. Accounting for this disruption effect, we can reproduce a drop in the fractional polarization of thermal dust emission with respect to dust temperature, above a critical value which depends on the tensile strength of the grains. The successful polarization model with RATD and a low tensile strength suggests a composite grain structure instead of a compact grain model, in agreement with \cite{2020ApJ...896...44L}.

We successfully reproduced the observed $P(\%)-T_{\rm d}$ trend in the case of $\rho$ Oph-A by considering both only silicate grains and mixed carbon and silicate grains to align with the magnetic field, assuming that the grain size distribution produced by the RATD follows a power-law distribution. With the parameters adapted in this work, our results indicate that composite grains with a power-index of size distribution steeper than the standard MRN distribution (i.e., $\beta<-3.5$) can reproduce the observational data, which well agrees with \cite{2015A&A...578A.131L}. Polarimetric data at longer wavelengths would help us to have a better understanding of grain alignment and disruption induced by RATs. In the forthcoming work, we combine these FIR data with 450 $\mu$m and 850 $\mu$m (\citealt{2018ApJ...859....4K}) data observed by JCMT to study the polarization spectrum.   

We thank the anonymous referee for helpful comments that improved the impact and the presentation of this paper. This research is based on observations made with the NASA/DLR Stratospheric Observatory for Infrared Astronomy (SOFIA). SOFIA is jointly operated by the Universities Space Research Association, Inc. (USRA), under NASA contract NNA17BF53C, and the Deutsches SOFIA Institut (DSI) under DLR contract 50 OK 0901 to the University of Stuttgart. Financial support for this work was provided by NASA through award 4$\_$0152 issued by USRA. T.H is funded by the National Research Foundation of
Korea (NRF) grants funded by the Korea government
(MSIT) through a Mid-career Research Program (2019R1A2C1087045). 
A.G is supported by the Programme National "Physique et
Chimie du Milieu Interstellaire" (PCMI) of CNRS/INSU with INC/INP co-funded by CEA and CNES. A.S. acknowledge support from the NSF through grant AST-1715876.

\end{document}